\documentclass[pra,amsmath,amssymb,twocolumn,superscriptaddress]{revtex4-2}
\usepackage{amsmath}
\usepackage{amssymb}
\usepackage{amstext}
\usepackage{amsfonts}
\usepackage{amsxtra}
\usepackage{bm}
\usepackage[usenames]{color}
\usepackage{grffile}
\usepackage{soul}
\usepackage{epstopdf}
\usepackage{xcolor}
\usepackage{graphicx}
\usepackage{marvosym} 
\usepackage{tikz}

\newcommand{\edit}[1]{{\color{black} #1}}
\newcommand\sbullet[1][.5]{\mathbin{\vcenter{\hbox{\scalebox{#1}{$\bullet$}}}}}

{%
\setlength{\fboxsep}{0pt}%
\setlength{\fboxrule}{1pt}%
}%

\begin{document}

\title{Alternating-domain supersolids in binary dipolar condensates}

\author{T. Bland}
 \affiliation{
     Institut f\"{u}r Quantenoptik und Quanteninformation, \"Osterreichische Akademie der Wissenschaften, Innsbruck, Austria
 }
 \affiliation{
     Institut f\"{u}r Experimentalphysik, Universit\"{a}t Innsbruck, Austria
 }
 
\author{E. Poli}
 \affiliation{
     Institut f\"{u}r Experimentalphysik, Universit\"{a}t Innsbruck, Austria
 }

\author{L. A. Pe\~na Ardila}
 \affiliation{
     Institut f\"{u}r Theoretische Physik, Leibniz Universit\"{a}t Hannover, Germany
 }

\author{L. Santos}
 \affiliation{
     Institut f\"{u}r Theoretische Physik, Leibniz Universit\"{a}t Hannover, Germany
 }
 
 \author{F. Ferlaino}
 \affiliation{
     Institut f\"{u}r Quantenoptik und Quanteninformation, \"Osterreichische Akademie der Wissenschaften, Innsbruck, Austria
 }
 \affiliation{
     Institut f\"{u}r Experimentalphysik, Universit\"{a}t Innsbruck, Austria
 }

\author{R. N. Bisset}
 \affiliation{
     Institut f\"{u}r Experimentalphysik, Universit\"{a}t Innsbruck, Austria
 }

\begin{abstract}
Two-component dipolar condensates are now experimentally producible, and we theoretically investigate the nature of supersolidity in this system.
We predict the existence of a binary supersolid state in which the two components form a series of alternating domains, producing an immiscible double supersolid.
Remarkably, we find that a dipolar component can even induce supersolidity in a nondipolar component.
In stark contrast to single-component dipolar supersolids, alternating-domain supersolids do not require quantum stabilization, and
the number of crystal sites is not strictly limited by the condensate populations, with the density hence being substantially lower.
Our results are applicable to a wide range of dipole moment combinations, marking an important step towards long-lived bulk-supersolidity.
\end{abstract}

\date{\today}
\maketitle

\section{Introduction}

The once elusive supersolid state of matter simultaneously exhibits superfluidity and crystalline order \cite{boninsegni2012colloquium}.
While early proposals sought superfluid properties of defects in a solid \cite{gross1957unified,andreev1969quantum}, focusing on helium experiments \cite{kim2004probable}, supersolidity has yet to be demonstrated in those systems \cite{Kim2012a}.
It is instead the high-degree of flexibility and control offered by ultracold gases that led to the first observations of supersolidity, but of a different kind, with solid properties arising in superfluids. Supersolid features were observed in systems with cavity-mediated interactions \cite{leonard2017supersolid},
while supersolid stripes were realized with spin-orbit coupled Bose-Einstein condensates (BECs) \cite{li2017stripe,Bersano2019}.
Supersolids have now been observed in experiments with dipolar BECs \cite{Tanzi2019,Bottcher2019,Chomaz2019}, and their superfluid character has been supported by the analysis of their excitations \cite{natale2019excitation,tanzi2019supersolid,guo2019low}.
Note that supersolid proposals have also been made for gases with soft-core, finite range interactions \cite{Cinti2010a,Saccani2011,Saccani2012,Kunimi2012,Macri2013}.

The first dipolar supersolids were realized in single-component BECs in cigar-shaped traps,
exhibiting a periodic density modulation along one direction \cite{Tanzi2019,Bottcher2019,Chomaz2019}, whereas experiments have now also created 2D supersolids with density modulations along two directions \cite{norcia2021two,Bland2022tds}.
\edit{From a theoretical perspective, there have been intriguing predictions for other exotic 2D supersolid states \cite{Bombin2017,Baillie2018,zhang2019supersolidity,zhang2021phases,hertkorn2021pattern,poli2021maintaining}, as well as alluring manifestations of quantum vortices \cite{gallemi2020quantized,roccuzzo2020rotating,ancilotto2021vortex}.}
Dipolar supersolids may be created from unmodulated BECs by inducing a roton instability \cite{Tanzi2019,Bottcher2019,Chomaz2019}.
Dipolar rotons---constituting a local minimum of the energy dispersion at finite momenta due to the anisotropic and long-ranged dipole-dipole interactions \cite{ODell2003a,santos2003roton}---were first observed in cigar-shaped \cite{Chomaz2018a,Petter2019} and then in pancake-shaped BECs \cite{schmidt2021roton}.
An unstable roton mode seeds a periodic density modulation that can subsequently be stabilized by quantum fluctuations as the density grows, resulting in a supersolid \cite{Boettcher2020,Chomaz2022}.
Recently, these same concepts were extended to the case of dipolar mixtures, i.e., systems composed of two dipolar components, which are now available in experiments \cite{Trautmann2018,Durastante2020,politi2022interspecies}.
In particular, it was predicted that exotic supersolid states can be seeded by the addition of a second dipolar component \cite{politi2022interspecies,scheierman2022catalyzation}.


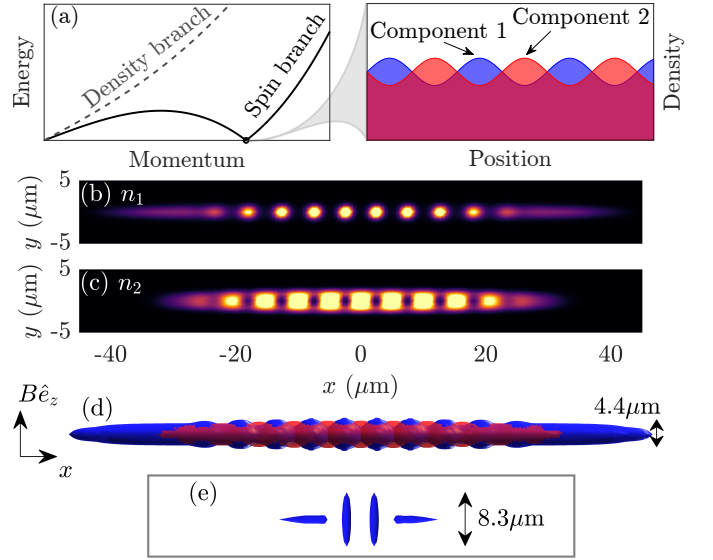
\begin{figure}[t]
    \centering
    \begin{tikzpicture}
        \node[anchor=south west,inner sep=0] at (-2.4,0) {\includegraphics[width=0.49\columnwidth]{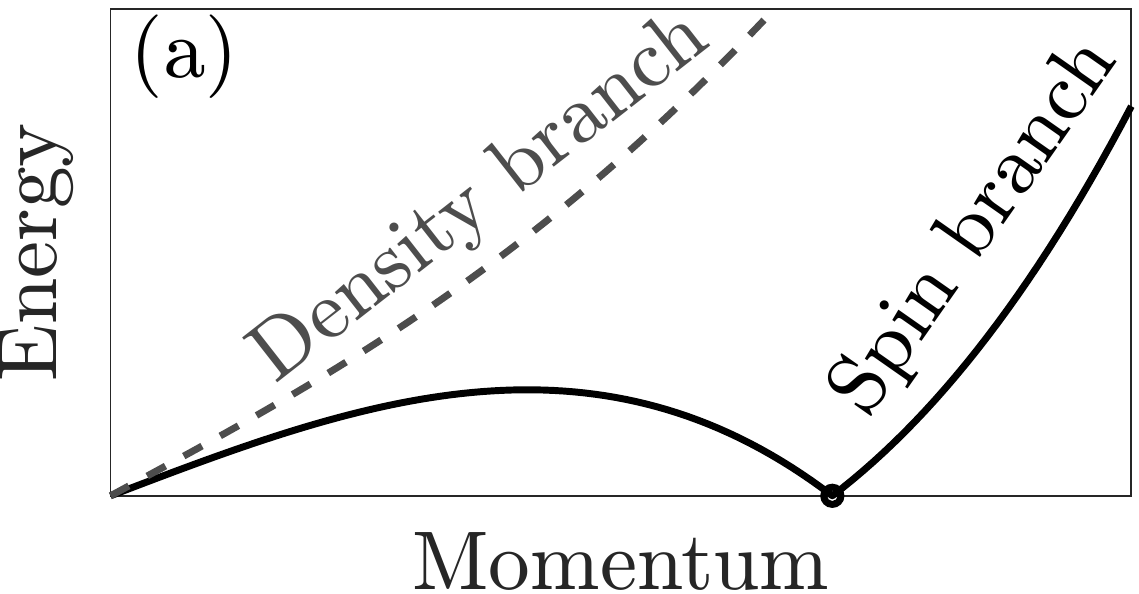}};
        \node[anchor=south west,inner sep=0] at (2.3,0) {\includegraphics[width=0.49\columnwidth]{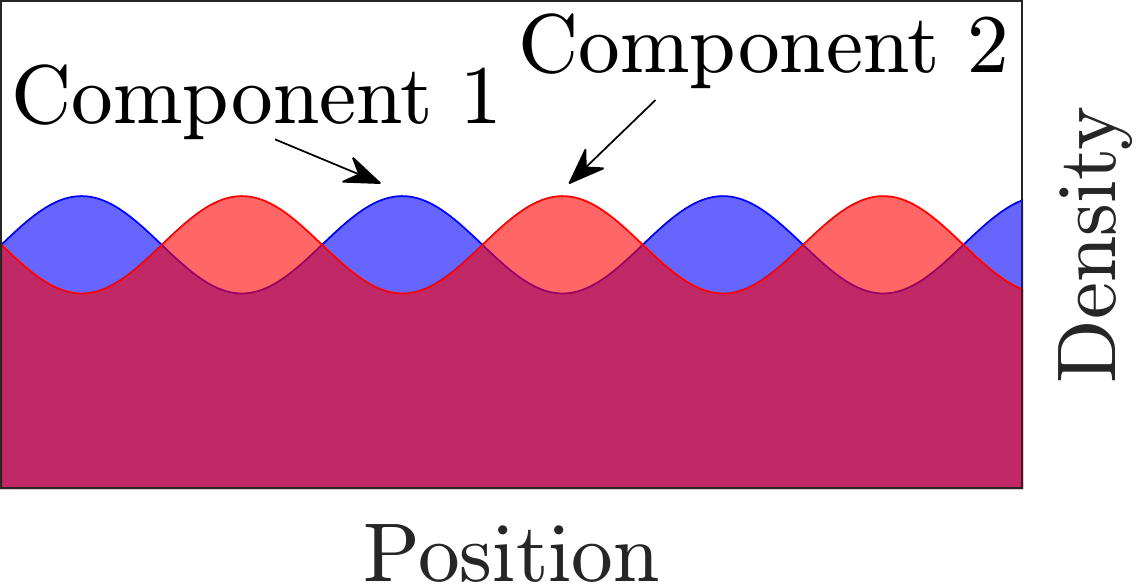}};
        \coordinate (A) at (0.7,0.36);
        \coordinate (B) at (2.3,0.39);
        \coordinate (C) at (2.3,2.2);
        \draw [fill,gray,opacity=0.2] (A) to[out=0,in=120] (B) -- (C) to[out=-100,in=0] (A);
        \draw [thick,gray,opacity=0.2] (A) to[out=0,in=120] (B) -- (C) to[out=-100,in=0] (A);
    \end{tikzpicture}
    \includegraphics[width=\columnwidth]{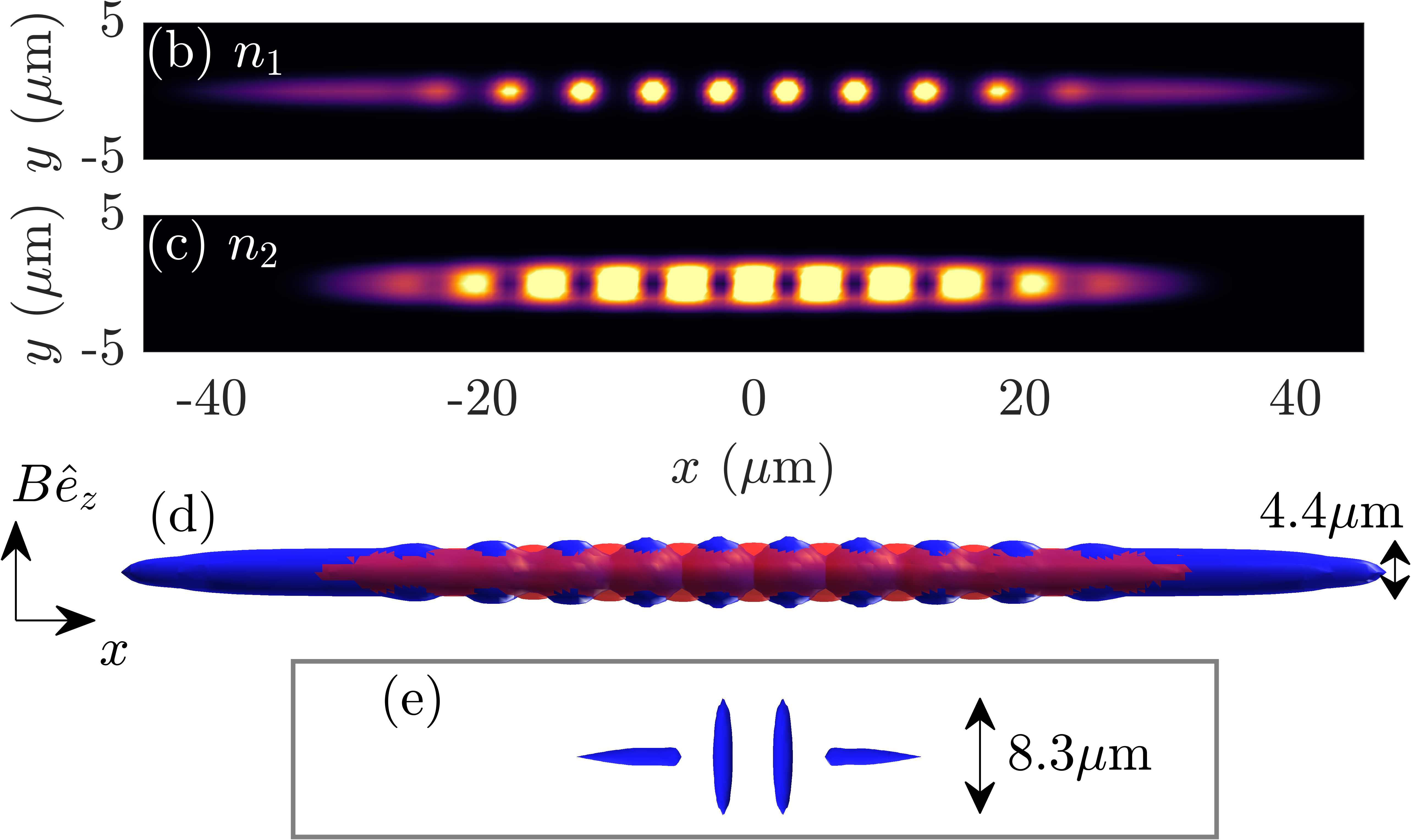}
    \caption{(a) Dispersion relation schematic showing a roton instability in the spin branch with corresponding density modulations on the right for a dipolar mixture. (b-d) Alternating-domain supersolid for a dysprosium dipolar-nondipolar mixture. Column densities for (b) dipolar and (c) nondipolar components, with (d) double-isosurface plot at 2\% of the peak density for each component. Interaction scattering lengths $(a_{11},a_{12},a_{22})=(100,98,100)a_0$, trapping frequencies $\vec{f}=(5,110,150)$Hz, and populations $N_1=N_2=1.5\times10^4$.
    (e) Corresponding single-component modulated state for the same trapping potential, $N=3\times10^4$ and $a = 78a_0$.
    Subplots (b-e) are drawn to the same length scale.}
    \label{fig:1}
\end{figure}


Dipolar mixtures, with their richness stemming from multiple sources of interactions, offer new phases with spontaneous modulation that go beyond quantum-fluctuation-stabilized supersolids.
An early example proposed that two immiscible BECs---displaced relative to one another by non-concentric confinement---might be used to realize a kind of binary supersolid formed by the instability of interface bending modes \cite{Saito2009}.
Binary dipolar BECs could also open another intriguing possibility.
It was already predicted that unmodulated binary BECs may be destabilized by a spin roton mode \cite{wilson2012roton,lee2021miscibility}, for which a periodic density modulation develops in both components, but with the density maxima of one component occurring at the minima of the other [see Fig.~\ref{fig:1}(a)].
The key question is then, in analogy to the dipolar roton producing a single-component supersolid, could the unstable spin rotons point to a novel kind of supersolid?

In this article, we predict the existence of a phase, which we call an alternating-domain supersolid, that exists even at the mean-field level and does not require the regulatory action of quantum fluctuations. The two components form alternating domains, with a continuous superfluid connection within each component that periodically weaves through the other [see Fig.~\ref{fig:1} (b-d)].
We uncover rich phase diagrams with broad regions in which both components are supersolid, as well as regions where a supersolid component is periodically punctuated by the isolated domains of the second component.
We predict that the alternating-domain supersolid intrinsically relies on a dipolar imbalance between the two components, and can exist for far lower atom numbers and peak densities than quantum-stabilized supersolids, which has important implications for the potential size and longevity of supersolid crystals in realistic settings.

\section{Formalism}
\edit{We consider a three-dimensional system at zero temperature made of two bosonic components, $\sigma=\{1,2\}$, consisting of atoms with permanent magnetic moments, although our work is also applicable to electric dipoles.
Following Refs.~\cite{Bisset2021,Smith2021}, we compute the Lee-Huang-Yang (LHY) energy density correction due to quantum fluctuations for a homogeneous binary mixture with densities $\mathbf{n}=(n_1,n_2)$:
\begin{align}
    \epsilon_\text{LHY}(\mathbf{n}) = \frac{16}{15\sqrt{2\pi}}\left(\frac{m}{4\pi\hbar^2}\right)^{3/2}\int_0^1\text{d}u\,\sum_{\lambda=\pm}V_\lambda(u,\mathbf{n})^{5/2}\,,
    \label{eq:LHYendens}
\end{align}
where we assume equal masses $m=m_1=m_2$, and
\begin{align}
    V_\pm(u,\mathbf{n}) = \sum_{\sigma=1,2}\alpha_{\sigma\sigma}n_\sigma\pm\sqrt{\left(\alpha_{11}n_1 - \alpha_{22}n_2\right)^2 + 4\alpha_{12}^2n_1n_2}\,.
\end{align}
Here, $\alpha_{\sigma\sigma'}(u) = g_{\sigma\sigma'} + g^d_{\sigma\sigma'}(3u^2-1)$, where the short-ranged and dipolar interaction parameters are, respectively, $g_{\sigma\sigma'} = 4\pi\hbar^2a_{\sigma\sigma'}/m$ and $g^d_{\sigma\sigma'} = \mu_0\mu_\sigma\mu_{\sigma'}/3 = 4\pi\hbar^2a^d_{\sigma\sigma'}/m$,
with $s$-wave scattering lengths, $a_{\sigma\sigma'}$ and dipole moments $\mu_\sigma$, where $\mu_0$ is the vacuum permeability.
The wave function for each component $\Psi_\sigma$ is obtained by solving the coupled extended Gross-Pitaevskii equations:
\begin{align} \label{eq:coupledGPE}
    i\hbar \frac{\partial}{\partial t} & \Psi_\sigma(\mathbf{x})  = \bigg[ -\frac{\hbar^2\nabla^2}{2m} + \frac{1}{2}m \left(\omega_x^2x^2+\omega_y^2 y^2 + \omega_z^2 z^2\right) \notag \\
    & + \sum_{\sigma'}\int \text{d}^3\mathbf{x}'\,U_{\sigma\sigma'}(\mathbf{x}^\prime-\mathbf{x})n_{\sigma'}(\mathbf{x}^\prime) + \sum_{\sigma'}g_{\sigma\sigma'}n_{\sigma'}(\mathbf{x}) \notag\\ 
    &+ \mu_{\text{LHY}}^{(\sigma)}[\mathbf{n}(\mathbf{x})] \bigg]\Psi_\sigma(\mathbf{x})\,,
\end{align}
where $\omega_{x,y,z}=2\pi f_{x,y,z}$ are the harmonic trapping frequencies, $U_{\sigma\sigma'}(\mathbf{r}) = \left[\mu_0\mu_\sigma\mu_{\sigma'}/4\pi r^3\right]\left(1-3\cos^2\theta\right)$ is the long-ranged anisotropic dipole-dipole interaction potential, with $\theta$ the angle between the polarization axis (always $z$) and the vector $\mathbf{r}$ that connects the two interacting particles, and $n_\sigma(\mathbf{x}) \equiv \left|\Psi_\sigma(\mathbf{x})\right|^2$ is the density of component $\sigma$, normalized to $N_\sigma$ atoms. The last term in (\ref{eq:coupledGPE}) is the quantum fluctuation correction to the chemical potential $\mu_{\text{LHY}}^{(\sigma)}[\mathbf{n}(\mathbf{x})] = \partial\epsilon_\text{LHY}(\mathbf{n}(\mathbf{x}))/\partial n_\sigma$, described within the local-density approximation framework.}

\section{Alternating-domain supersolids}
We demonstrate the unique features of alternating-domain supersolids by considering a dipolar-nondipolar mixture in Fig.~\ref{fig:1}, for which a combined total of 19 domains can be seen.
While the dipolar component [Fig.~\ref{fig:1}(b)] can remain globally phase coherent through a continuous superfluid connection linking the domains---since we are close to the miscible-immiscible transition the separation is only partial---the nondipolar component [Fig.~\ref{fig:1}(c)] can also maintain a superfluid connection along high density rails encompassing the dipolar domains.
The density isosurfaces in Fig.~\ref{fig:1}(d) highlight the shape of the dipolar domains,
which are not as strongly elongated as the single-component case [cf.~Fig.~\ref{fig:1}(e)].
While this concrete illustration considers two $^{164}$Dy spin projections, with $(\mu_1,\mu_2)=(-10,0)\mu_B$ for Bohr magneton $\mu_B$, domain supersolids are not just a special feature of dipolar-nondipolar mixtures, but are rather general, as we discuss later.

These results must be contrasted to the single-component case. In Fig.~\ref{fig:1}(e) we show a modulated state for a single-component dipolar BEC for the same trap and total atom number as in Fig.~\ref{fig:1}(b-d), i.e., $N=3\times10^4$.
Note that we had to modify the scattering length, since $a=100a_0$ corresponds to an unmodulated BEC. However, lowering to $a=78a_0$
\edit{passes a transition to a modulated state that is not a supersolid}, where the peak density ($4.0\times10^{21}$m$^{-3}$) is immediately more than an order of magnitude larger than the domain supersolid case ($1.8\times10^{20}$m$^{-3}$). Correspondingly, the number of atoms per lattice site is about an order of magnitude larger than for the domain supersolid. For this atom number and trap volume, the supersolid phase does not exist for the single-component case \cite{poli2021maintaining}, which was also the situation for the regimes considered by Refs.~\cite{Kadau2016a,Wachtler2016a,Bisset2016}.

To understand the physical mechanisms involved, it is instructive to consider the transition from unmodulated to modulated states. For the formation of a domain supersolid, the density modulation is triggered by unstable spin roton excitations [shown schematically in Fig.~\ref{fig:1}(a)], with wavelengths governed by the BEC’s width along the direction of dipole polarization \cite{wilson2012roton}. \edit{Spin modes act to increase the density difference}, $|n_1 - n_2|$, and the instability is hence resolved once the components become spatially separated as alternating immiscible domains [[Figs.~\ref{fig:1}(b-d)]. \edit{Crucially, there is no significant increase of the total density $n_1 + n_2$}, and the peak density can remain low. This situation should be contrasted to that of single component supersolids\edit{, and the two-component supersolids of Refs.\,\cite{politi2022interspecies,scheierman2022catalyzation},} for which \edit{a density} roton instability would cause a divergence of the peak density if it were not counterbalanced by the appropriate LHY term \cite{Lima2011a}, and this necessitates significantly higher densities \cite{Tanzi2019,Bottcher2019,Chomaz2019}. \edit{Note that while single component supersolids require $a < a^d$, and thus are only stable due to quantum fluctuations \cite{Boettcher2020,Chomaz2022}, domain supersolids can exist for either $a_{\sigma\sigma} > a^d_{\sigma\sigma}$ or $a_{\sigma\sigma} < a^d_{\sigma\sigma}$, but quantum fluctuations remain qualitatively important since, for example, if they were neglected the latter situation could only be at best metastable \footnote{\edit{If quantum fluctuations are neglected, $a < a^d$ generally means that dipolar condensates can only be metastable \cite{Lahaye_RepProgPhys_2009}}}.}

We focus on regimes where one component without the presence of the other will always be unmodulated,
but each component within the binary system can exist in one of three phases: an unmodulated BEC, a supersolid state with a linear chain of domains (SS), or an array of isolated domains (ID). The distinction between these is set by upper-bound estimates for the superfluid fractions \cite{leggett1970can}, which in our binary system are given by
\begin{align*}
    f_{s,\sigma} = \frac{(2L)^2}{\int_{-L}^L\text{d}x\, \bar{n}_\sigma}\left[\int\displaylimits_{-L}^L\frac{\text{d}x}{\bar{n}_\sigma}\right]^{-1},~ \bar{n}_\sigma=\int\displaylimits _{-\infty}^\infty\int\displaylimits_{-\infty}^\infty\text{d}y\text{d}z\,n_\sigma\,,
\end{align*}
for length $L$ defined over the central region that encompasses the central 3 (4) domains if the number of domains is odd (even).
We take the supersolid region to be when $f_{s,\sigma}>0.1$ occurs concurrently with a periodic density modulation, \edit{following Ref.\,\cite{blakie2020supersolidity} for the arbitrary choice of $f_{s,\sigma}\leq0.1$ defining the crossover to the regime of isolated domains}.
For reference, the superfluid fractions in Fig.~\ref{fig:1} are (b) $f_{s,1} = 0.3032$, (c) $f_{s,2} = 0.7940$ and (e) $f_{s,1} = 0.0001$. Note, the total superfluid fraction $f_s = (N_1f_{s,1} + N_2f_{s,2})/N$ is associated with a reduction in the moment of inertia of the overall mixture. The periodic spatial ordering can be characterized by the density contrast ${\mathcal{C}_\sigma = \left(n^\text{max}_\sigma - n^\text{min}_\sigma\right)/\left(n^\text{max}_\sigma + n^\text{min}_\sigma\right)}$, where $n^\text{max}_\sigma$ ($n^\text{min}_\sigma$) are neighboring maxima (minima) as one moves along the trap's long direction. \edit{See Appendix \ref{app:contrast} for more discussion on the calculation of the contrast.}
The boundary between unmodulated and modulated states is defined by $\mathcal{C}_\sigma$ changing from zero to a nonzero value.

\begin{figure}[t]
    \centering
    \includegraphics[width=1\columnwidth]{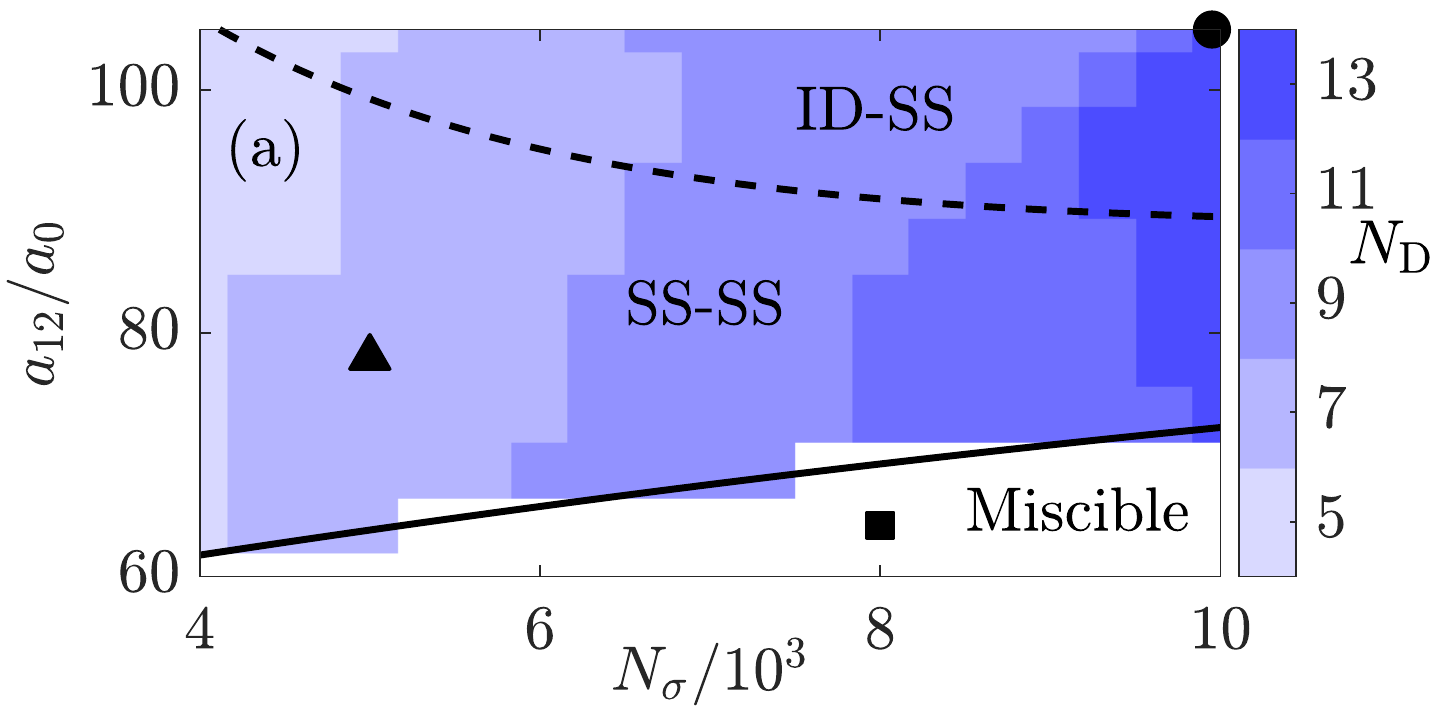}\vspace{-0.3cm}\\
    \includegraphics[width=0.41\columnwidth]{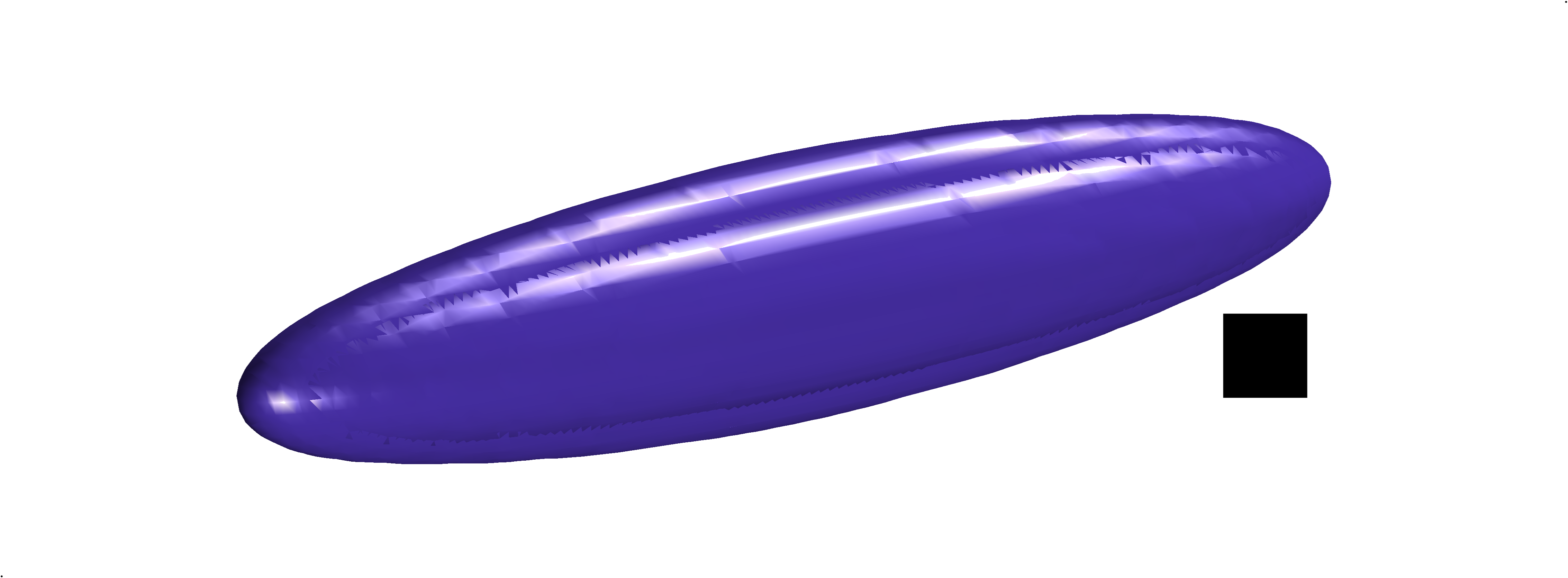}\hspace{-1.3cm}
    \includegraphics[width=0.41\columnwidth]{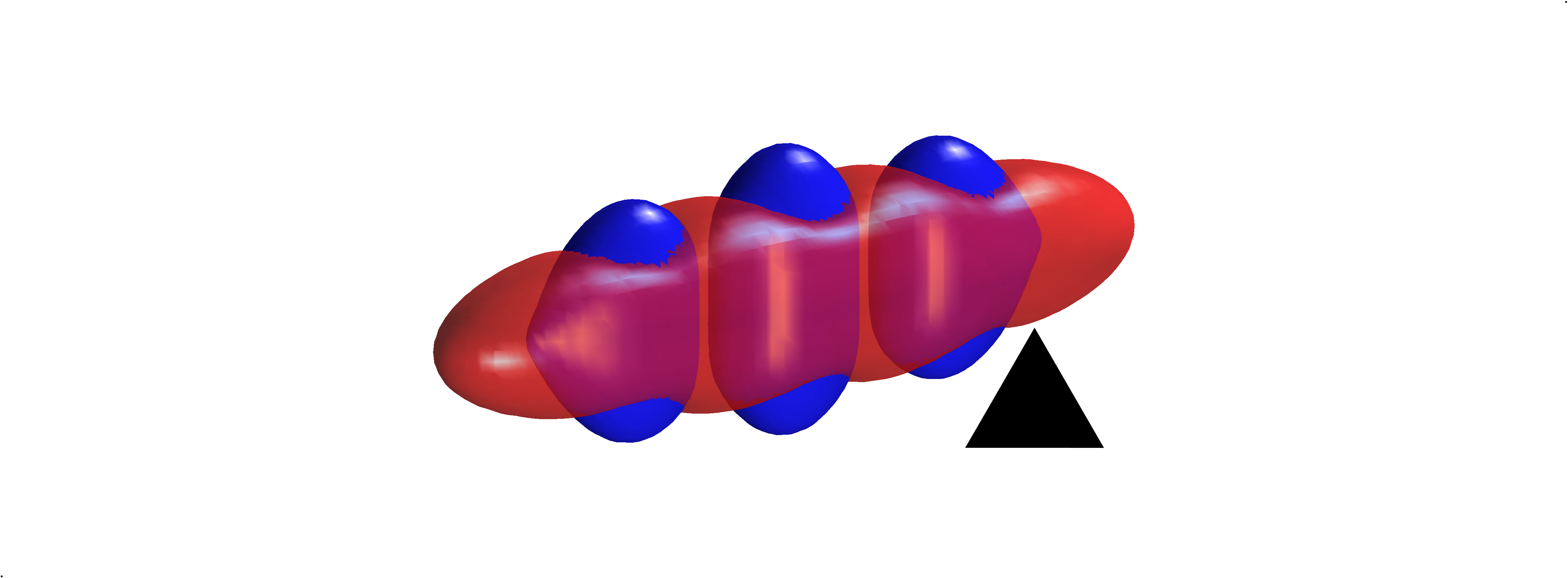}\hspace{-1.2cm}
    \includegraphics[width=0.41\columnwidth]{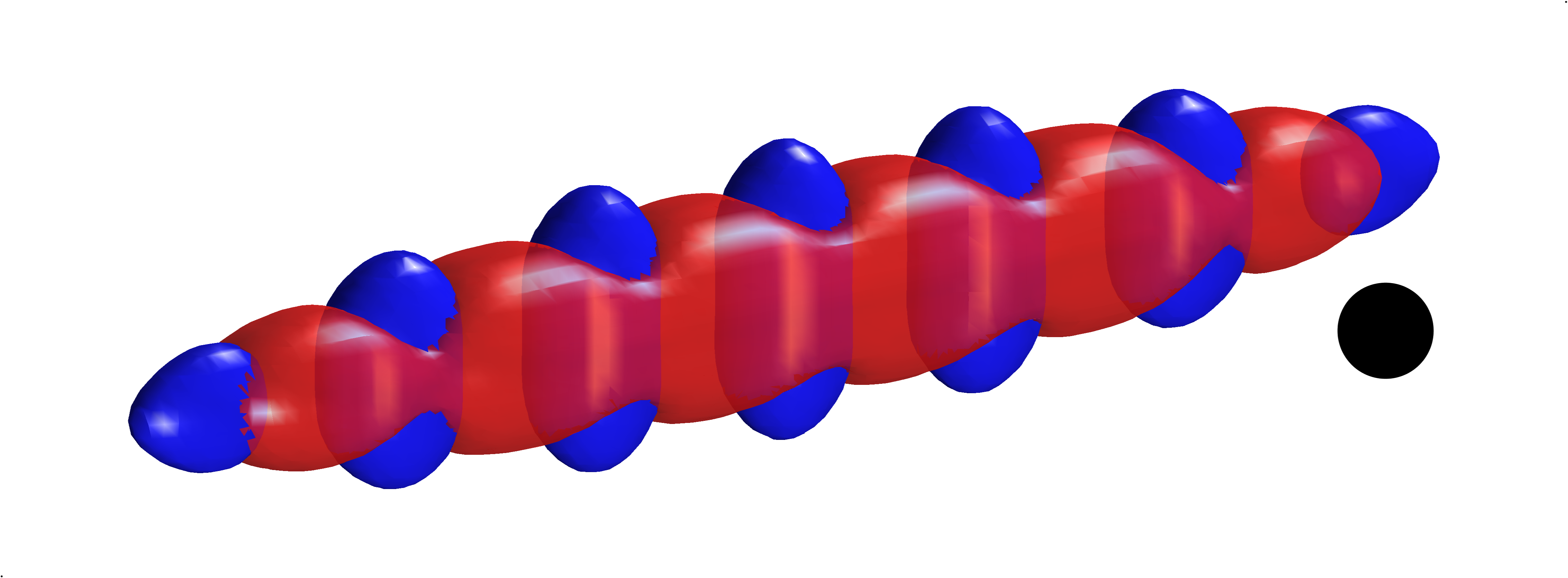}\vspace{-0.1cm}\\
    \includegraphics[width=0.495\columnwidth]{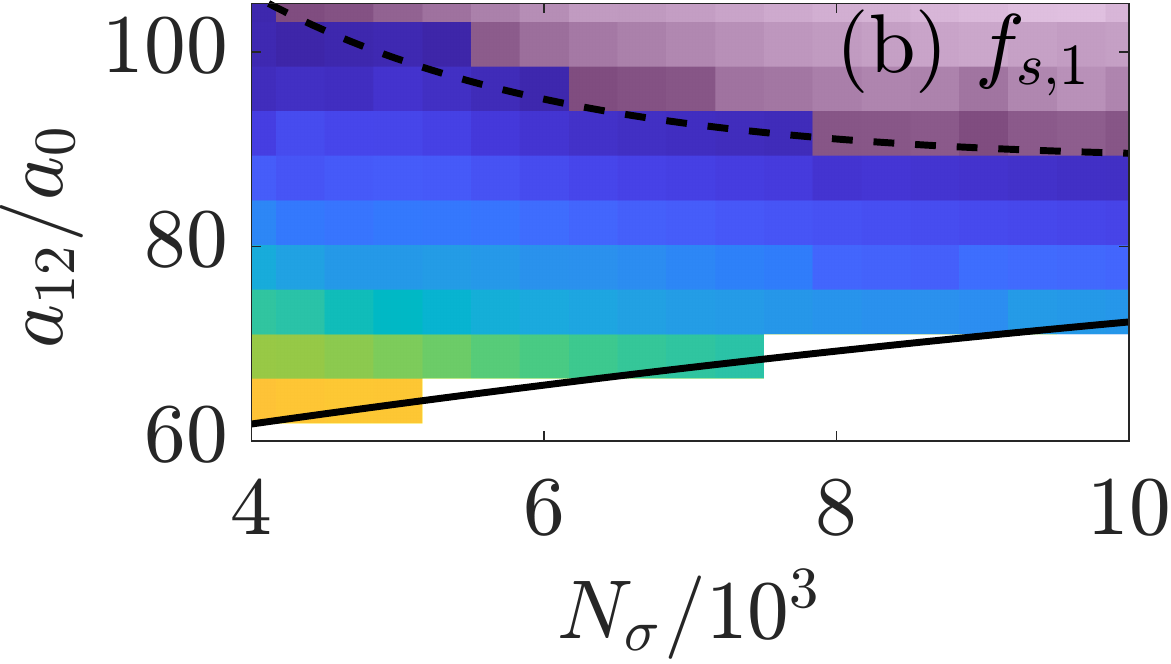}
    \includegraphics[width=0.4875\columnwidth]{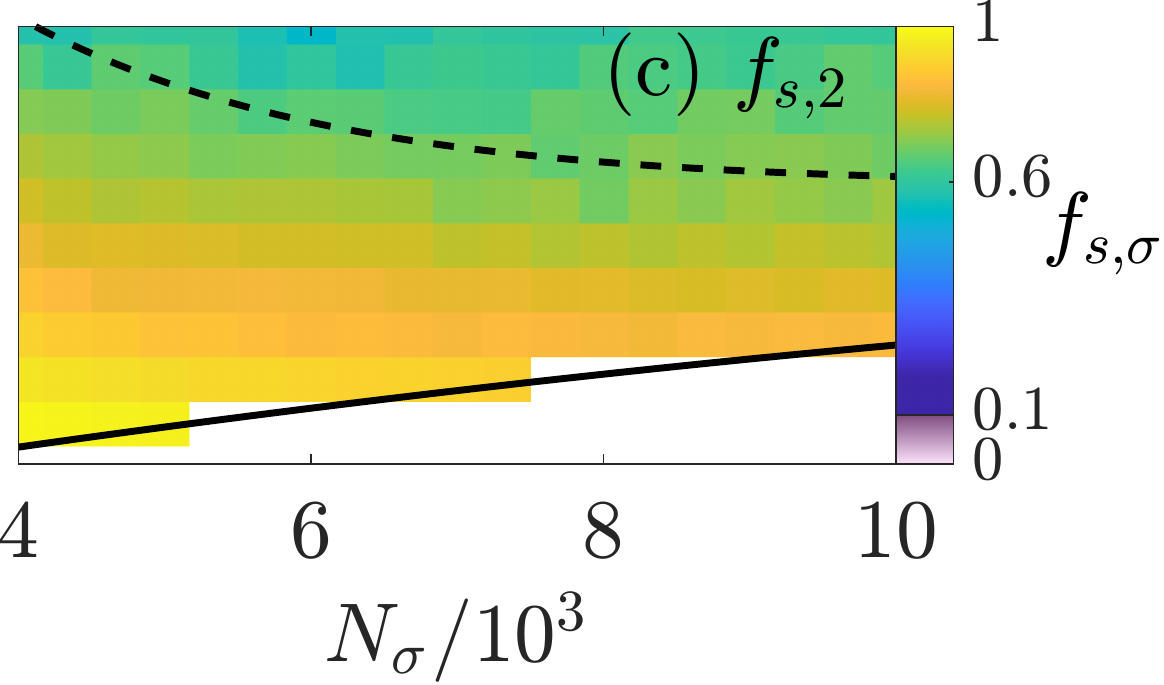}
    \caption{Phase diagram for a dipolar-nondipolar mixture of $^{164}$Dy atoms, varying intercomponent scattering length $a_{12}$ and $N_1 = N_2 = N/2$ with fixed $Nf_x$, from $f_x=37.5$Hz on the left to $f_x=15$Hz on the right. (a) Total number of domains in the stationary state solution. Solid lines separate unmodulated miscible to domain supersolid (SS-SS) state when $\mathcal{C}_\sigma>0$, and dashed line to isolated domains-supersolid (ID-SS) state. Example isosurfaces below are highlighted by the symbols in (a). (b,c) Superfluid fraction $f_{s,\sigma}$ of components 1 and 2. Threshold between SS and ID regime is indicated by a change of color scale. Other parameters: $a_{11}=a_{22}=100a_0$, $(\mu_1,\mu_2)=(-10,0)\mu_B$, $f_y=f_z=150$Hz.}
    \label{fig:2}
\end{figure}

\section{Phase diagram for dipolar-nondipolar mixture}
In Fig.~\ref{fig:2} we explore the stationary state phase diagram of a dipolar-nondipolar mixture in a cigar-shaped trap with ($f_x$,150,150)Hz, $N_1=N_2=N/2$ and fixed $Nf_x=3\times10^5$Hz to maintain an approximately constant average density \cite{poli2021maintaining}. At low $a_{12}\lesssim60a_0$ the stationary state solution is a miscible unmodulated BEC, with only small deviation from perfect density overlap between components due to magnetostriction in the dipolar component [Fig.~\ref{fig:2}$\,\blacksquare$]. Increasing $a_{12}$ induces a transition to a domain supersolid state (SS-SS) [Fig.~\ref{fig:2}$\,\blacktriangle$], where the domains of a given component exhibit a continuous superfluid connection [$f_{s,\sigma}>0.1$ in Figs.~\ref{fig:2}(b)(c)].
We find that a quench of the intercomponent scattering length from the unmodulated miscible state to the domain supersolid regime generates a globally phase-coherent state---within each component---that is robust against the excitations induced by the quench, in-keeping with single-component studies of supersolids in a cigar-shaped geometry \cite{Tanzi2019,Bottcher2019,Chomaz2019}, \edit{which we detail in Appendix \ref{app:dynamic} \cite{video1}}. Note how broad the SS-SS regime is, at least 20$a_0$ wide, compared to single-component supersolids where it is typically only a few $a_0$ wide \cite{blakie2020supersolidity}. Further increasing $a_{12}$ causes the overlap between components to reduce, expanding the distance between domains whilst {\color{black}decreasing superfluidity} [Fig.~\ref{fig:2}(b)], crossing into the isolated domain-supersolid (ID-SS) regime [Fig.~\ref{fig:2}$\,\sbullet[1.5]$].
However, the nondipolar component maintains a strong superfluid connection [Fig.~\ref{fig:2}(c)].
Note that the superfluid connection of component 2 can be controlled by adjusting $f_y$, with even a small reduction in $f_y$ significantly reinforcing the nondipolar rails around the dipolar domains.

Figure \ref{fig:2}(a) also shows how the total number of domains $N_{\rm D}$ changes in this phase diagram. Throughout, the average atom number per domain is $\sim10^3$.
Hence, as the atom number increases the number of domains climbs steadily, reaching a total of 13 once the system has $2\times10^4$ atoms ($10^4$ per component) on the far right-hand side.
In contrast, {\color{black}single-component dipolar supersolids} typically require $\sim10^4$ atoms per {\color{black} lattice site} \cite{Tanzi2019,Bottcher2019,Chomaz2019}.

\begin{figure}
    \centering
    \includegraphics[width=1\columnwidth]{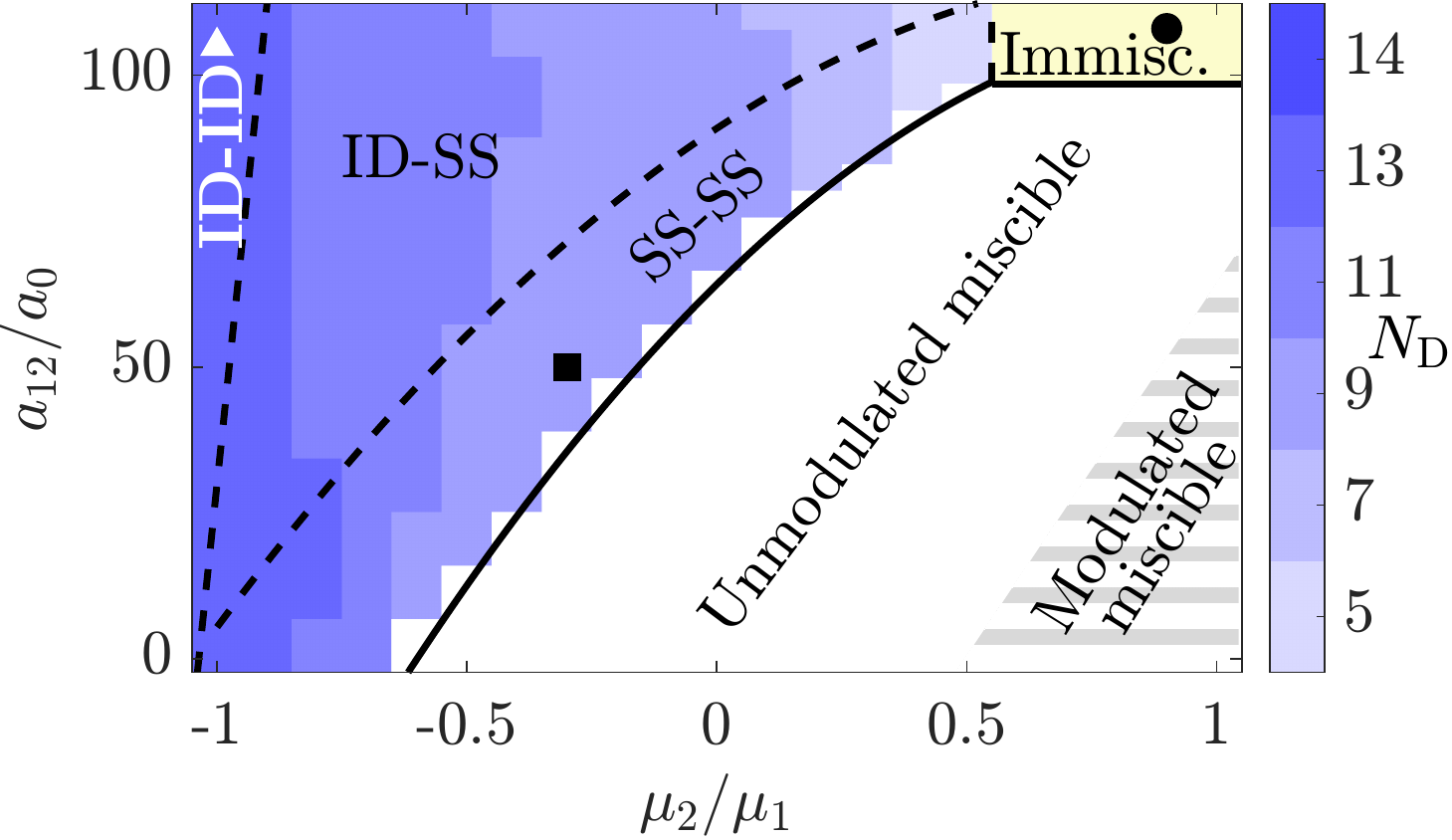}\vspace{-0.2cm}\\
         \hspace{-1.1cm}\includegraphics[width=0.43\columnwidth]{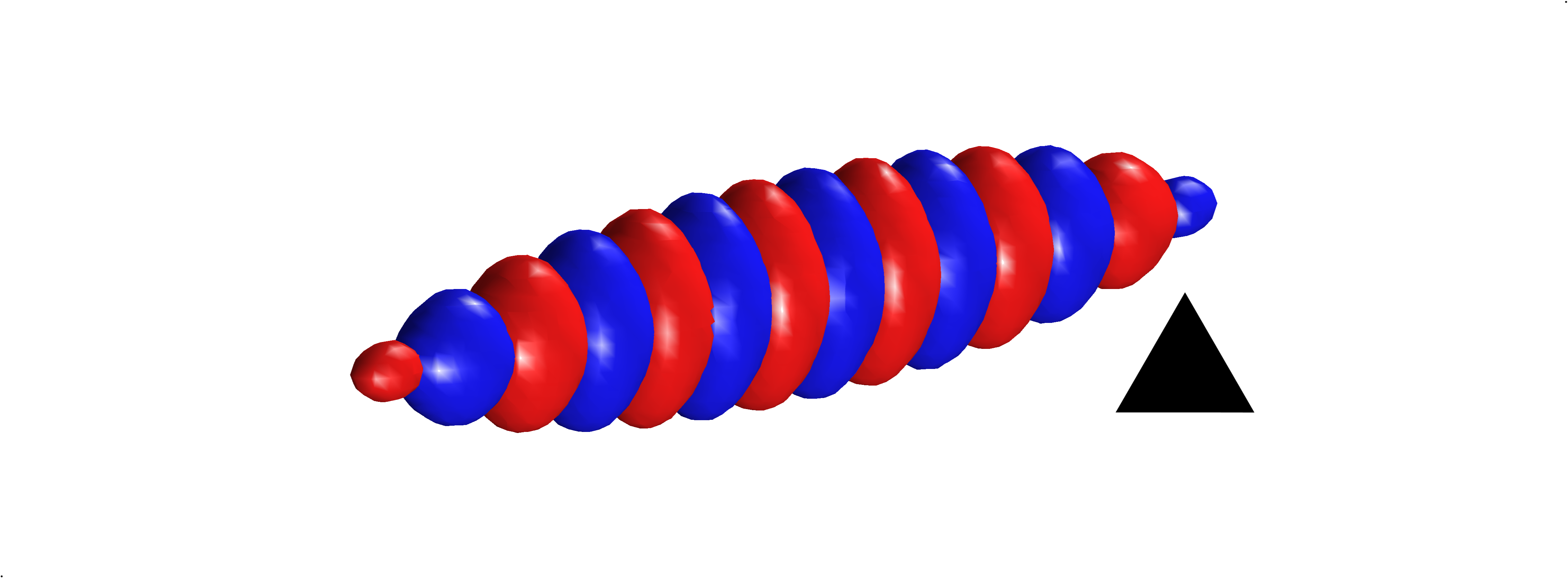}\hspace{-1.1cm}
    \includegraphics[width=0.43\columnwidth]{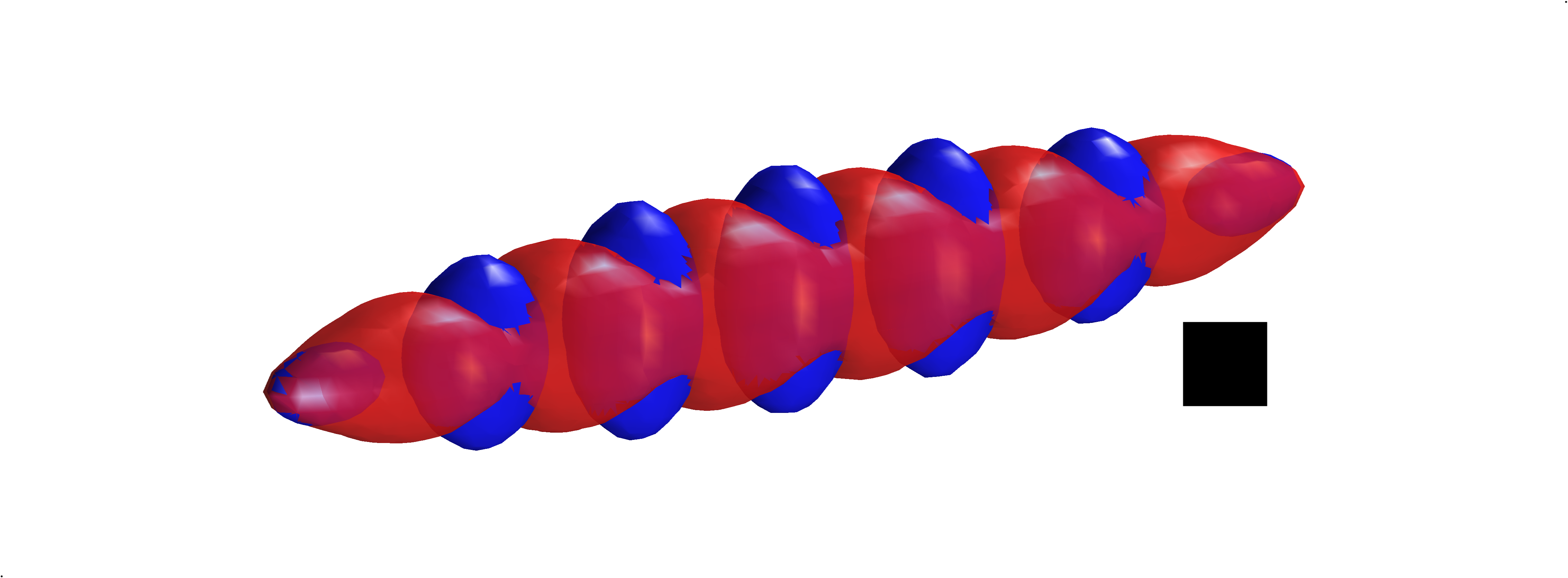}\hspace{-0.8cm}
    \includegraphics[width=0.43\columnwidth]{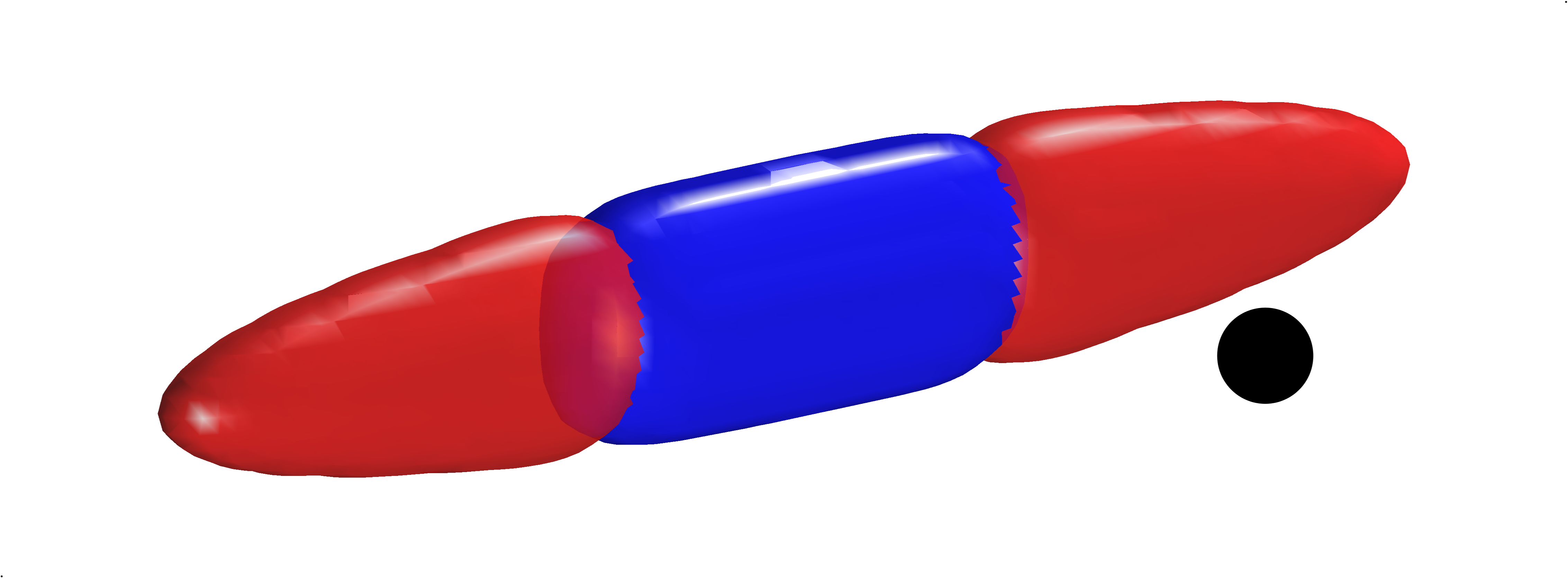}\vspace{-0.4cm}
    \caption{Phase diagram for dipolar mixtures with varying inter-component scattering length and relative magnetic moment (note that $\mu_2/\mu_1<0$ implies antiparallel dipoles). Compared to Fig.~\ref{fig:2}, note the new phases: binary isolated domains (ID-ID), macroscopic-domain immiscibility (beige region), and the modulated miscible regime.
    Parameters: $N_1 = N_2 = 5\times10^3$, $a_{11}=a_{22}=100a_0$, and $(f_x,f_y,f_z) = (15,150,150)$Hz.}
    \label{fig:3}
\end{figure}

\section{Generalization to various dipole combinations}
Here, we generalize our findings to mixtures in which both components can be dipolar, applicable to a wide range of experiments, e.g., erbium-dysprosium mixtures or spin mixtures of the same species.
In Fig.~\ref{fig:3}, we construct a phase diagram by fixing $\mu_1$ and exploring the effect of varying $\mu_2$ and $a_{12}$.
A solid line indicates a transition from a miscible to immiscible state, consistent with Fig.~\ref{fig:2}.
For $\mu_2/\mu_1<0$ the dipoles are anti-aligned,
decreasing the energy for dipoles of separate components to orient in the side-by-side configuration, thus causing both immiscibility and domain supersolidity to occur at low $a_{12}$ [Fig.~\ref{fig:3}$\,\blacksquare$].
At $\mu_2/\mu_1=-1$ the modulation is a perfect reflection about the $x=0$ plane between the components, and for the range of $a_{12}$ considered the system forms a binary isolated domain (ID-ID) state with 14 domains for only $10^4$ atoms in total [Fig.~\ref{fig:3}$\,\blacktriangle$]. 

For similar dipoles $\mu_2\sim\mu_1$ there is little energy incentive from the dipolar interactions for the components to phase separate \cite{lee2021miscibility}, hence the immiscibility boundary in Fig.~\ref{fig:3} is close to the nondipolar result $a_{12}=\sqrt{a_{11}a_{22}}=100a_0$, and the components separate to a macroscopic-domain immiscible state \cite{wilson2012roton} [Fig.~\ref{fig:3}$\,\sbullet[1.5]$].
While we focus on the immiscible domain regime, smaller $a_{\sigma\sigma}$ can trigger the formation of immiscible quantum-stabilized supersolids \cite{politi2022interspecies}.
Miscible quantum-stabilized supersolids are also possible for smaller $a_{12}$ \edit{following a density branch roton instability}, indicated in the lower right corner of Fig.~\ref{fig:3}, which is explored further in Ref.~\cite{scheierman2022catalyzation}. \edit{If we instead considered $a_{\sigma\sigma} \geq a^d_{\sigma\sigma}$ we would expect qualitatively the same phases as in Fig.~\ref{fig:3} apart for the modulated miscible phase.}

\begin{figure}[t]
    \centering
    \includegraphics[width=1\columnwidth]{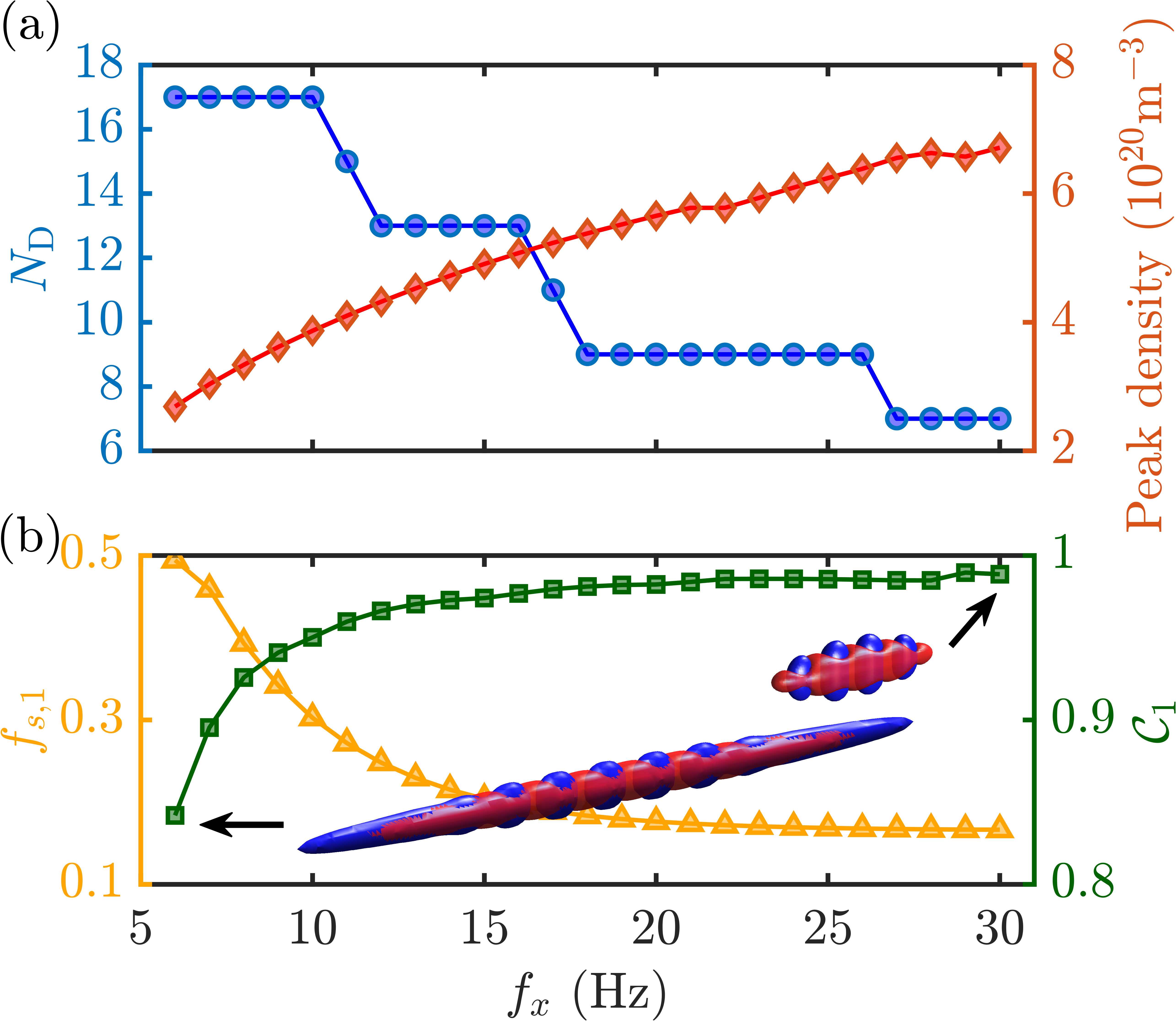}
    \caption{Opening the trap for an erbium dipolar-nondipolar supersolid. (a) Reducing the long axis trap frequency, $f_x$, of a cigar-shaped trap increases the number of domains $N_{\rm D}$ whilst simultaneously reducing the peak density. (b) Superfluid fraction (density contrast, $\mathcal{C}_1$) of the dipolar component also increases (decreases), indicating an improving superfluid connection, whilst the second component is always a robust supersolid with the superfluid fraction never dropping below 0.7 (not shown). Parameters: $\mu_1 = -7\mu_B$ and $\mu_2 = 0$, $(a_{11},a_{12},a_{22}) = (65,60,65)a_0$, $N_1 = N_2 = 20000$ atoms, $f_y=f_z = 150$\,Hz.}
    \label{fig:4}
\end{figure}

\section{Ultra-low density supersolids}
We investigate weakening the axial confinement of a dipolar-nondipolar spin mixture of erbium, further demonstrating the generality of our results to a broad range of dipole combinations.
On the far right of Fig.~\ref{fig:4} is a state in the SS-SS regime for $f_x=30$Hz.
Decreasing $f_x$ to 6Hz increases the total number of domains from 7 to 17, whilst simultaneously reducing the peak density by a factor of $\approx 2.5$.
The increasing number of domains can be explained by the BEC becoming longer, while the spin roton wavelength is roughly fixed by the confinement length in the direction of dipole polarization.
This behavior starkly contrasts with that for {\color{black} quantum-stabilized} supersolids, which instead require a certain atom number for a given trap volume \cite{poli2021maintaining}, and the supersolid regime is not possible if this criterion is not met \cite{Kadau2016a,Wachtler2016a,Bisset2016}. For example, {\color{black} recall the modulated state} in Fig.~\ref{fig:1}(e), for which the atom number is insufficient for this trap to attain supersolidity. Whilst decreasing $f_x$ the superfluid fraction is monotonically increased from close to the ID-SS to deep in the SS-SS regime. These results are compared to the density contrast, which shows an improved density linking between domains (smaller $\mathcal{C}_\sigma$) for looser confinement.

\section{Conclusions}
We predict an alternating-domain supersolid state in two-component dipolar condensates.
This binary supersolid exists over a broad region of parameter space and, importantly, it is robust against the excitations caused by crossing the unmodulated BEC--to--domain supersolid transition.
There is also a crossover to an adjacent region where one of the components is supersolid, but the other forms isolated domains. In contrast to single-component supersolids--which must be stabilized by quantum fluctuations--alternating-domain supersolids can produce numerous lattice sites with relatively small atom numbers, and have similar peak densities to unmodulated BECs, important for their longevity, which is largely determined by the inelastic three-body collisions that depend strongly on the density \cite{Chomaz2022}.

Our results are applicable to various dipole moment combinations, such as spin mixtures or binary gases comprised of two atomic species. Interestingly, we even find that a dipolar component can induce supersolidity within a non-dipolar component via their mutual interactions.
Our work opens the door for future investigations into binary supersolid states and their excitations, as well as the exploration of novel 2D domain supersolids with exotic structures and vortex states. Our results reveal a rich system, within current experimental reach, and mark an important step towards long-lived bulk-supersolidity.

{\it Note added}.---Very recently, we became aware of a simultaneously submitted work addressing supersolidity in an immiscible dipolar-nondipolar mixture \cite{li2022long}. \\

\section*{Acknowledgements}
We thank Danny Baillie, P.~Blair Blakie, and Wyatt Kirkby for stimulating discussions. Part of the computational results presented here have been achieved using the HPC infrastructure LEO of the University of Innsbruck. T.~B.~acknowledges funding from FWF Grant No. I4426 2019. We acknowledge support of the Deutsche Forschungsgemeinschaft (DFG, German Research Foundation) under Germany’s Excellence Strategy – EXC-2123 QuantumFrontiers – 390837967. R.~B. acknowledges financial support by the ESQ Discovery programme (Erwin Schrödinger Center for Quantum Science \& Technology), hosted by the Austrian Academy of Sciences (ÖAW).

\appendix

\section{Density contrast} \label{app:contrast}

\begin{figure}
    \centering
    \includegraphics[width=1\columnwidth]{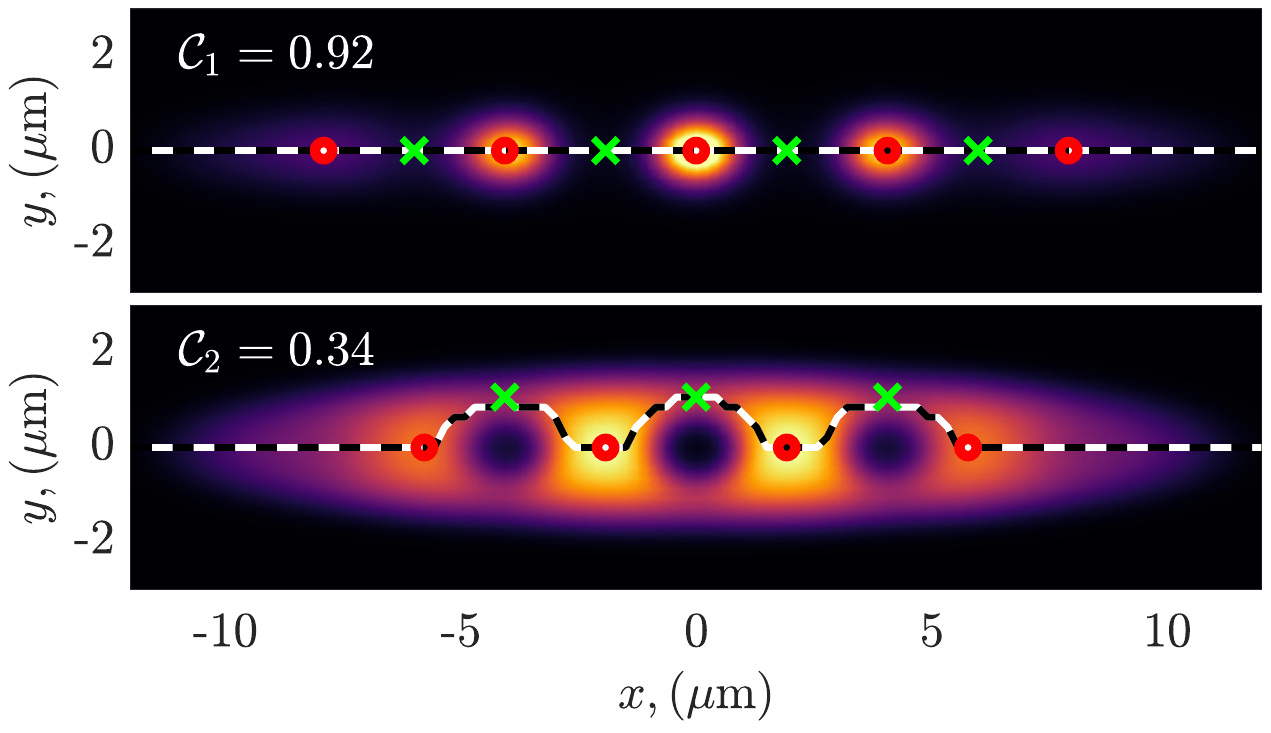}
    \caption{Contrast $\mathcal{C}_\sigma$ of a dipolar-nondipolar mixture of $^{164}$Dy atoms. Dashed line indicates the $y$ position of the maximum density along $x$. Red circles are the peaks ($n^\text{max}_\sigma$) and green crosses the troughs ($n^\text{min}_\sigma$) of density along the dashed line. Other parameters: $(\mu_1,\mu_2)=(-10,0)\mu_B$, $N_1=N_2=N/2=7000$, $a_{12}=70a_0$, $a_{11}=a_{22}=100a_0$, $(f_x,f_y,f_z)=(21,150,150)$Hz.}
    \label{fig:S1}
\end{figure}

The onset of periodic density modulation is characterized by the density contrast, akin to interferometric visibility, defined as
\begin{align}
    \mathcal{C}_\sigma = \frac{n^\text{max}_\sigma - n^\text{min}_\sigma}{n^\text{max}_\sigma + n^\text{min}_\sigma}
    \label{eqn:Csig}
\end{align}
for each bosonic component $\sigma = \{1,2\}$ and where $n^\text{max}_\sigma$ and $n^\text{min}_\sigma$ are neighboring maxima and minima in the 3D density as one moves along the long direction of the trap. In Fig.~\ref{fig:S1} we graphically depict the line of maximum 3D density in the $z=0$ plane, showing the maxima (red circles) and minima (green crosses) in the density along this curve. Typically, for the component with a larger dipole moment this curve lies along $y=0$, just as it does for single-component supersolids \cite{Tanzi2019,Bottcher2019,Chomaz2019}.
However, the nondipolar component has a greater superfluid connection along the rails, a feature which can be captured by our generalization of Eq.~\eqref{eqn:Csig}. This connection can still be lost, however, through tightening $f_y$, for example.

\section{Dynamic preparation} \label{app:dynamic}

\begin{figure*}
\begin{tikzpicture}
    \node[anchor=south west,inner sep=0] at (-1,0) {\includegraphics[width=0.32\linewidth]{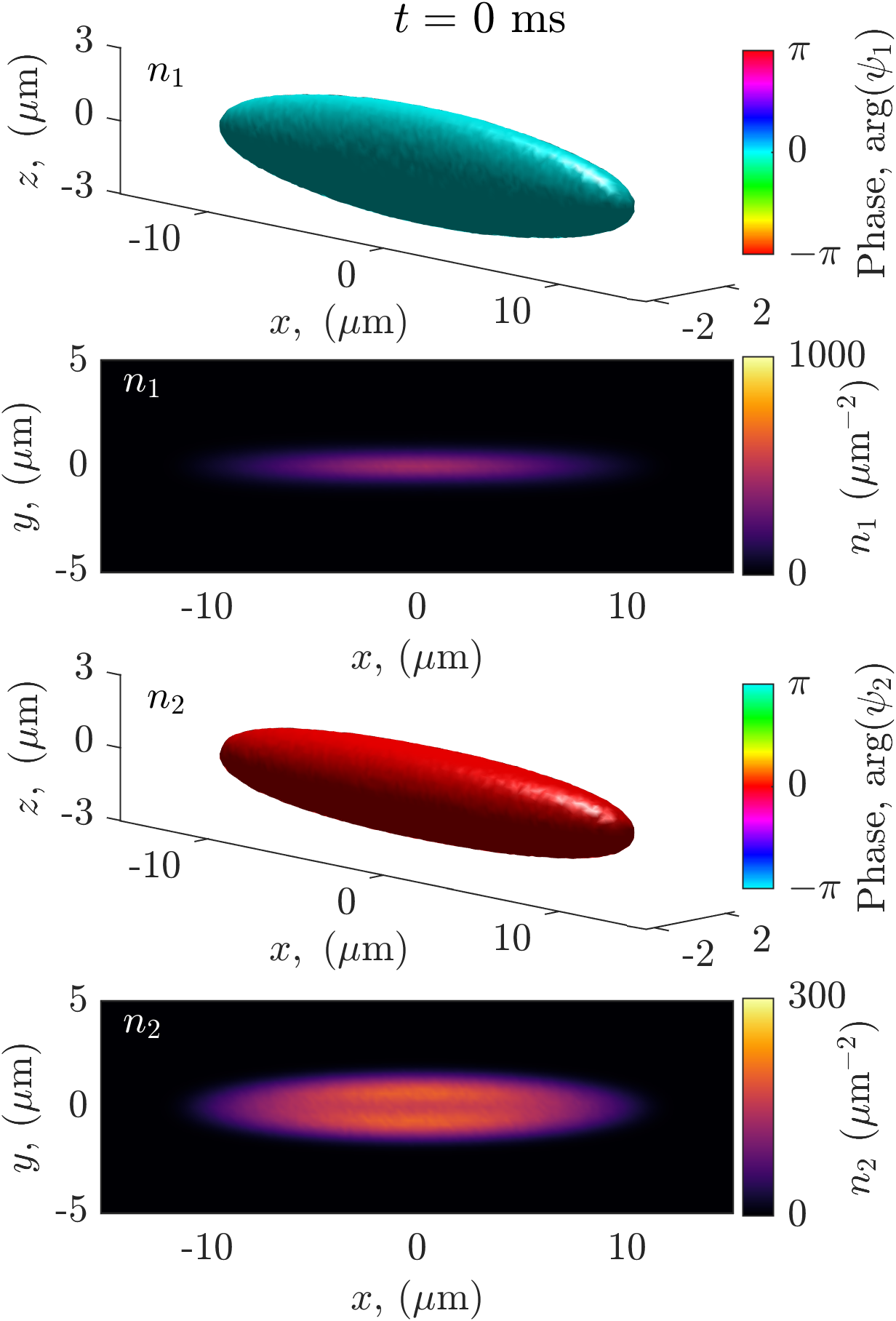}};
    \node[anchor=south west,inner sep=0] at (5,0) {\includegraphics[width=0.32\linewidth]{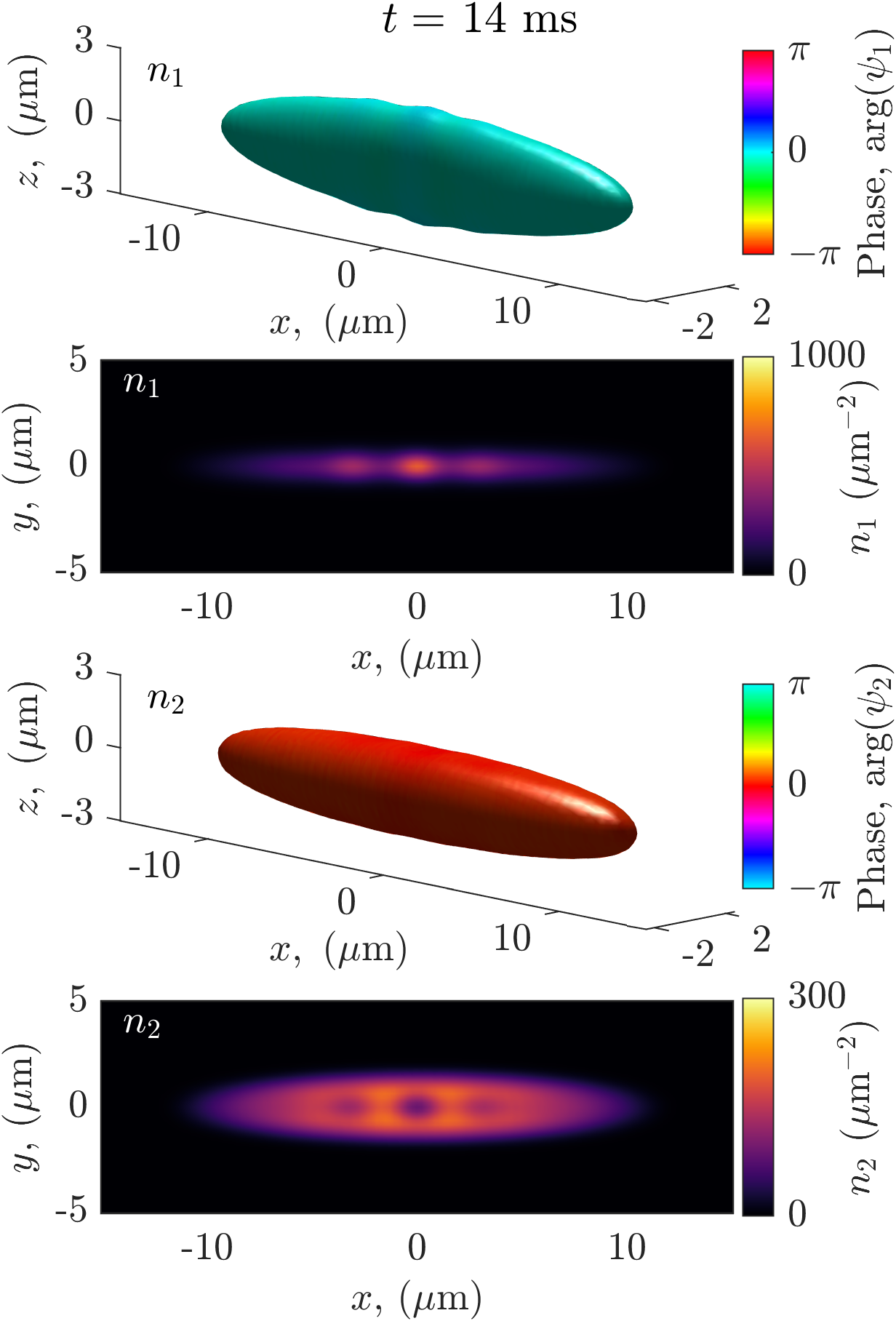}};
    \node[anchor=south west,inner sep=0] at (11,0) {\includegraphics[width=0.32\linewidth]{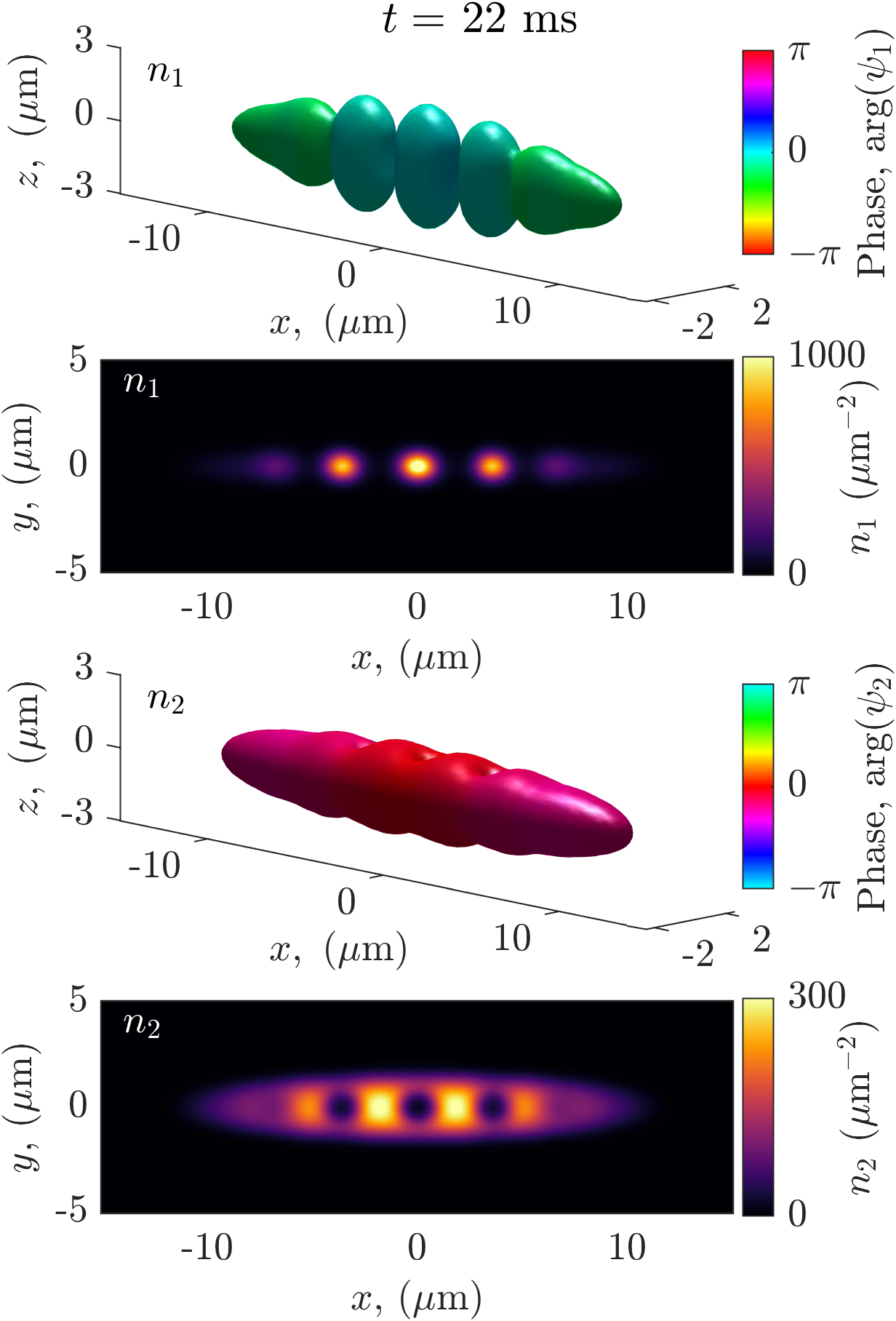}};
     \draw[black, thick] (4.9,0) -- (4.9,8.5);
     \draw[black, thick] (10.9,0) -- (10.9,8.5);
     \node[anchor=south west,inner sep=0] at (-1,-9) {\includegraphics[width=0.32\linewidth]{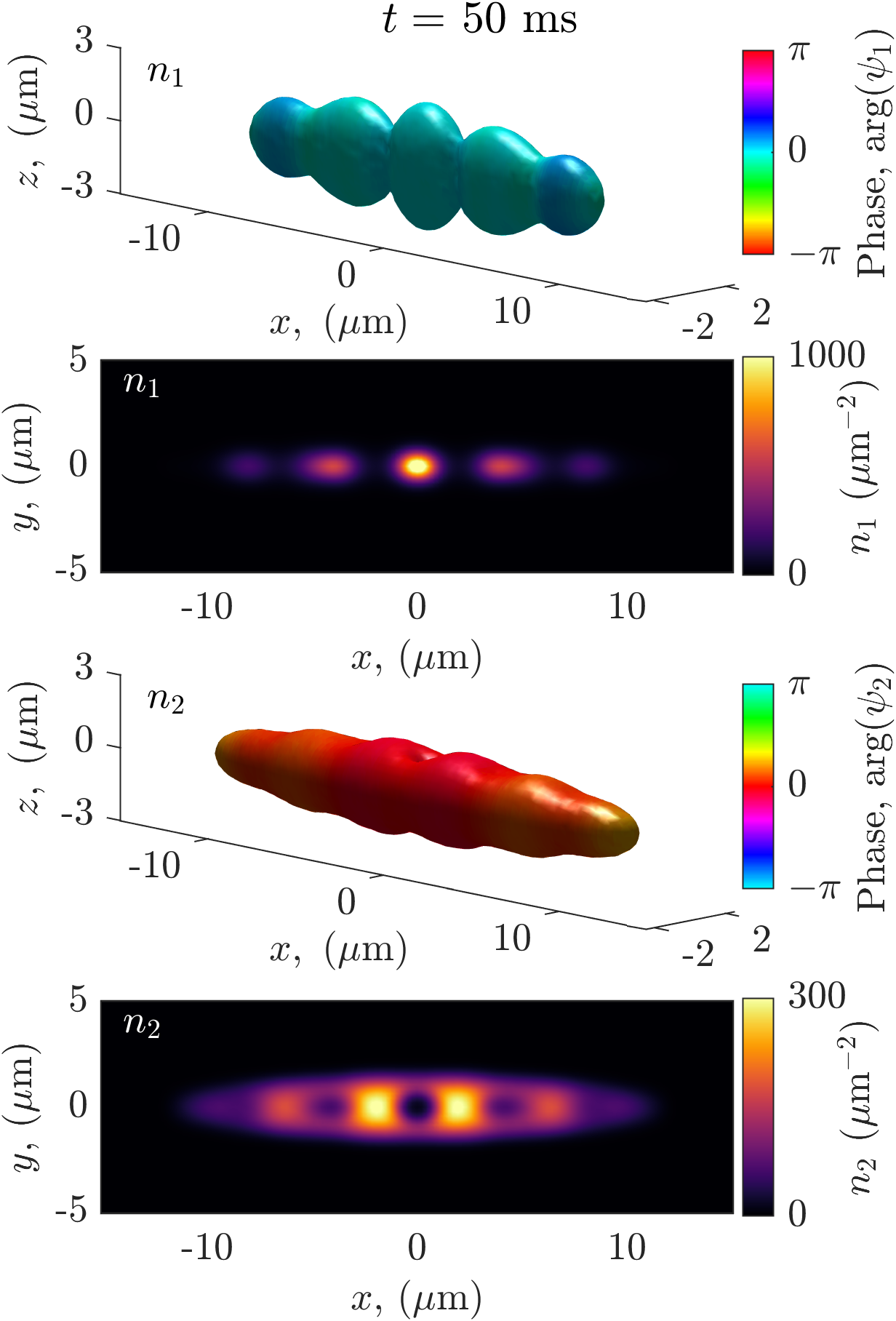}};
    \node[anchor=south west,inner sep=0] at (5,-9) {\includegraphics[width=0.32\linewidth]{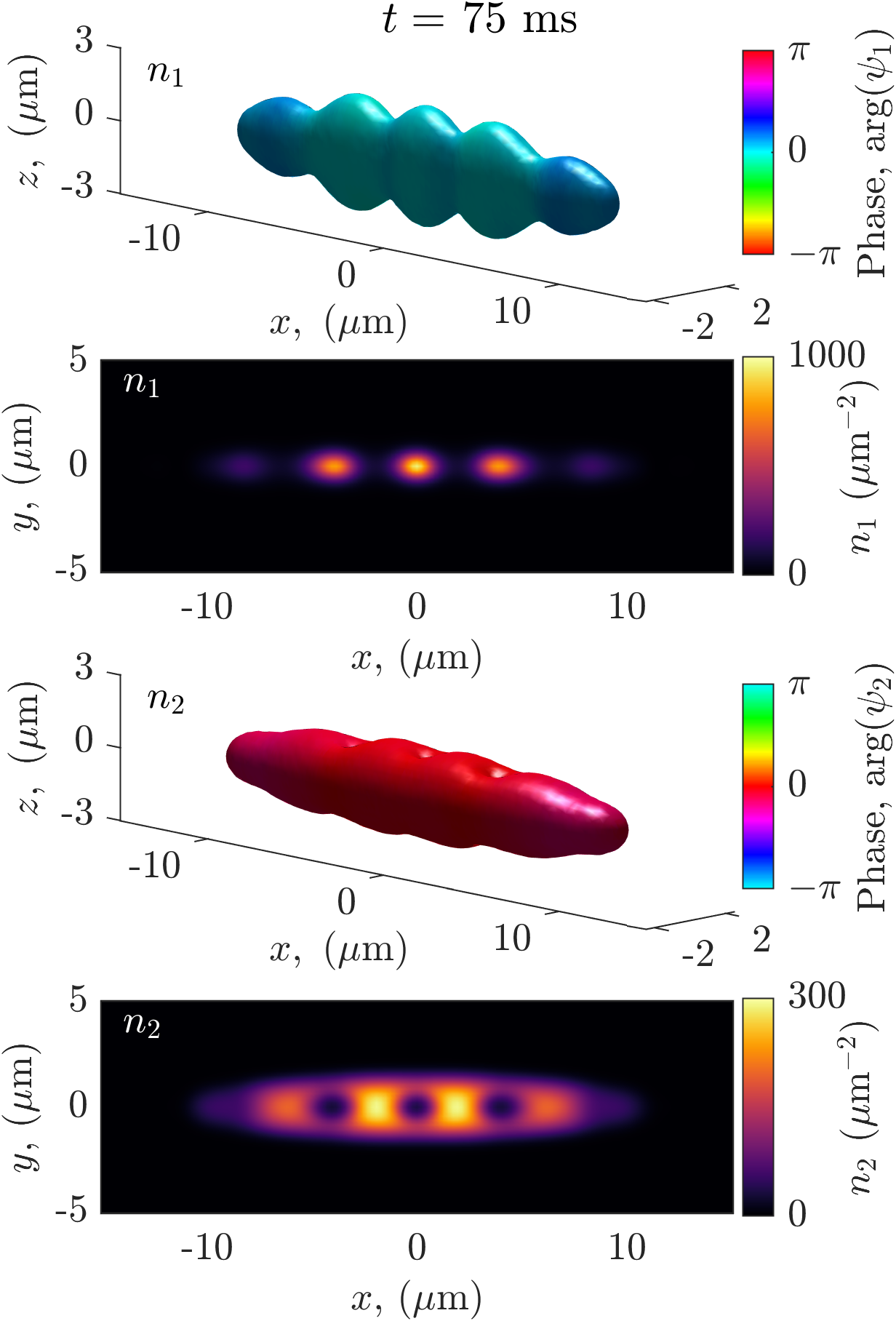}};
    \node[anchor=south west,inner sep=0] at (11,-9) {\includegraphics[width=0.32\linewidth]{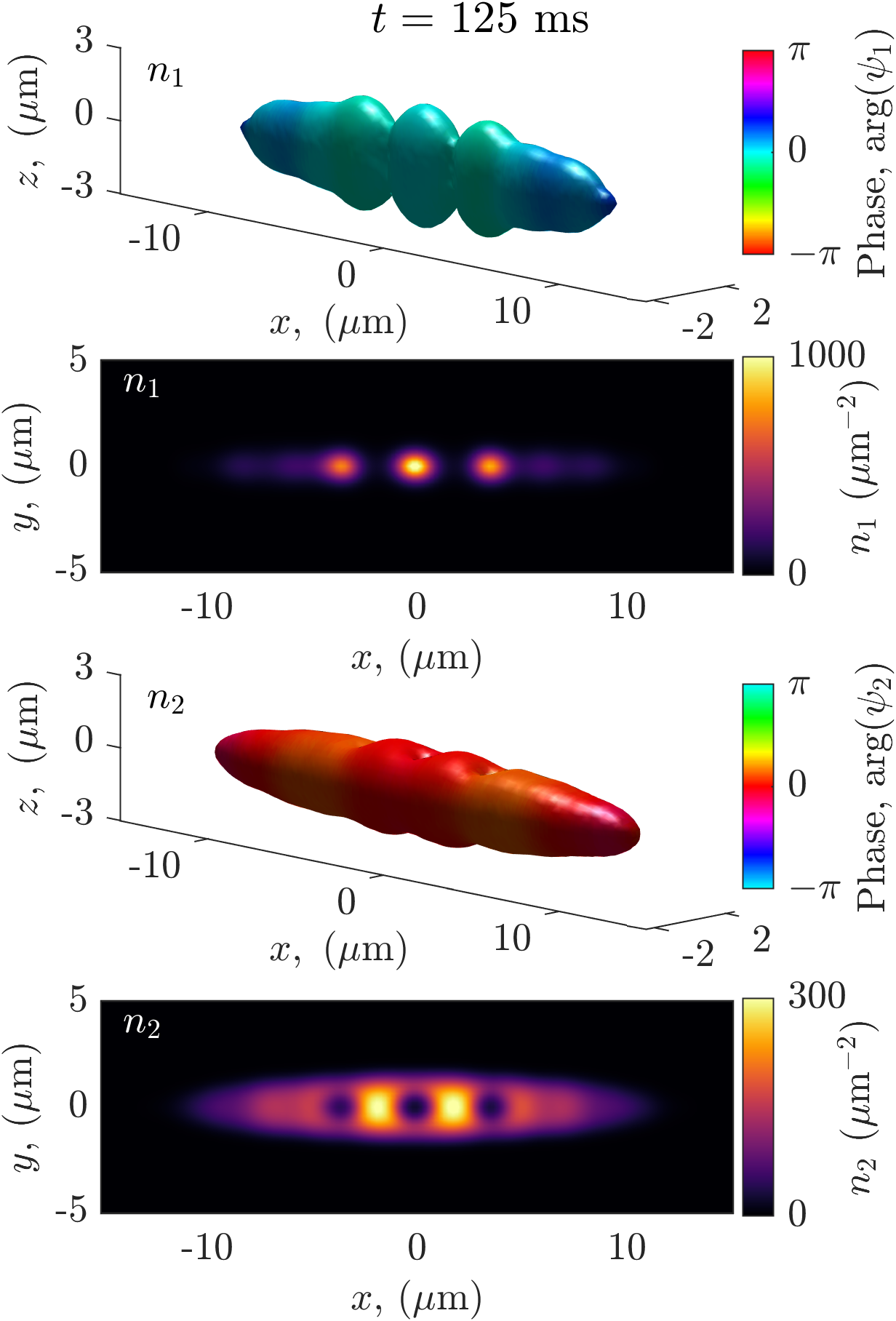}};
    \draw[black, thick] (4.9,0) -- (4.9,-9);
    \draw[black, thick] (10.9,0) -- (10.9,-9);
    \draw[black, thick] (-1,-0.3) -- (16.6,-0.3);
    \node at (1.2,8.35) {(a)};
    \node at (7.1,8.35) {(b)};
    \node at (13.1,8.35) {(c)};
    \node at (1.15,-0.66) {(d)};
    \node at (7.1,-0.66) {(e)};
    \node at (13.1,-0.66) {(f)};
\end{tikzpicture}
  \caption{Preparation of a domain supersolid through an interaction quench. Simulation of an instantaneous quench from $a_{12}=65a_0$ to $a_{12}=70a_0$ at $t=0$, with other parameters from Fig.~\ref{fig:S1}. Each labeled frame corresponds to a time during the consequent dynamics, and within each frame the data can be understood row-by-row. Row 1: 5\% density isosurface for component 1 colored to the phase, and centered such that the phase at the origin is 0, perfect coherence for component 1 would be light blue. Row 2: Column density for component 1 normalized to peak value over the whole simulation. Row 3: same as Row 1 but for the second component, but perfect coherence in component 2 would be red. Row 4: same as Row 2 but for the second component, with a smaller peak density. Note that the stationary solution for the final parameters is the state presented in Fig.~\ref{fig:S1}. A Supplementary Video of this simulation is also included \cite{video1}.}
  \label{fig:S2}
\end{figure*}
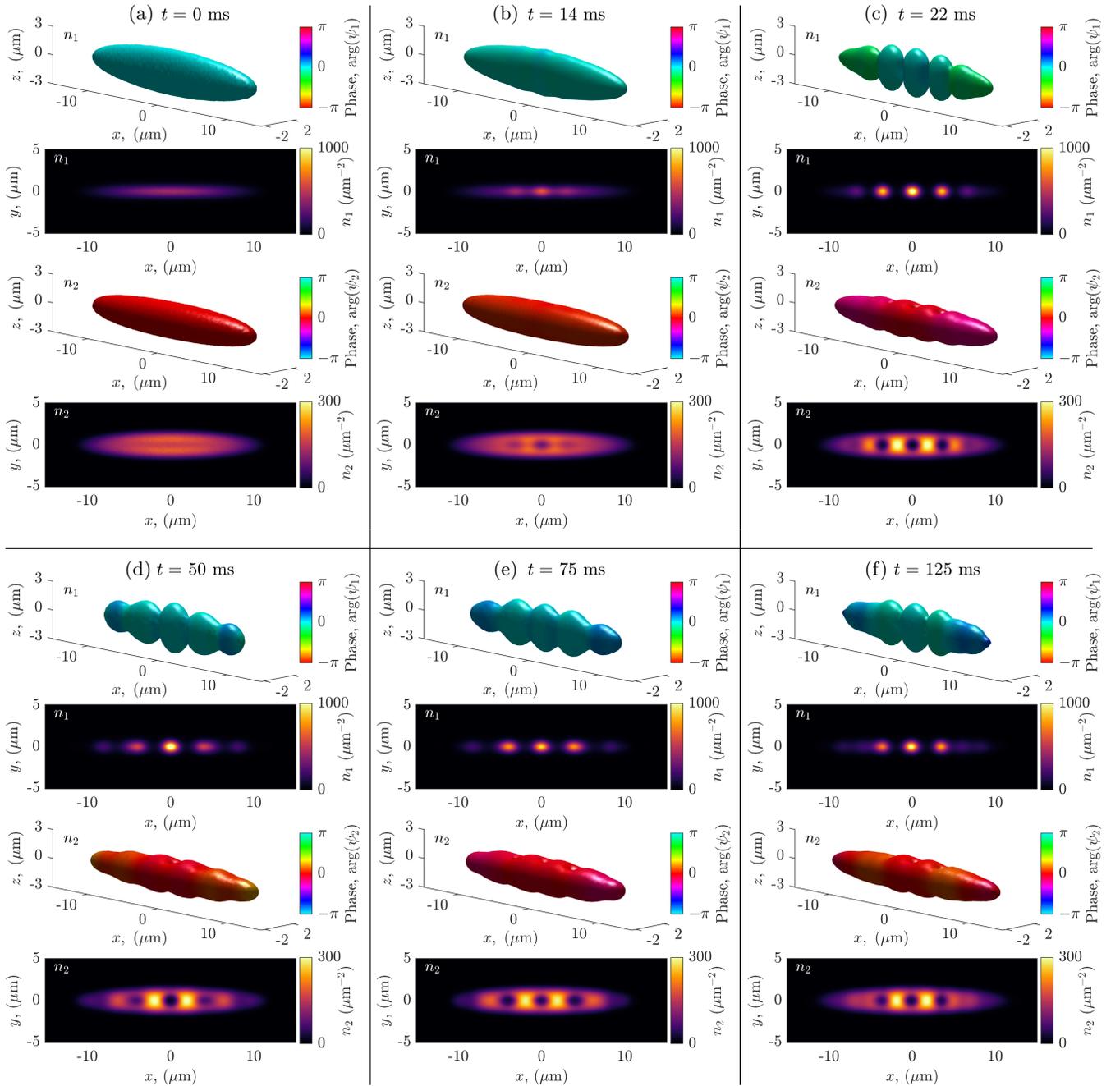

The preparation of a single-component supersolid has been achieved through either taking an unmodulated BEC and quenching the scattering length across the unmodulated BEC-to-supersolid transition \cite{Tanzi2019,Bottcher2019,Chomaz2019}, or by direct evaporative cooling into the supersolid state \cite{Chomaz2019,norcia2021two,Bland2022tds}. The two-component case affords a wide range of possibilities for domain supersolid preparation, due to the large number of tunable interaction parameters in the system. Here, we investigate one possibility through tuning the intercomponent scattering length $a_{12}$. Taking an initially unmodulated miscible dipolar-nondipolar mixture with the parameters from Fig.~\ref{fig:S1} and $a_{12}=65a_0$, we simulate an instantaneous quench to $a_{12}=70a_0$. The consequent dynamics are shown in Fig.~\ref{fig:S2}. Despite the violent nature of the instantaneous quench, the system maintains phase coherence throughout the lifetime of the simulation, as indicated by the blue (red) isosurface for component 1 (2), and the solution resembles the target stationary solution [Fig.~\ref{fig:S1}]. We also include a Supplementary Video of the dynamics \cite{video1}.


\clearpage

\begin{figure}
      \centering
      \includegraphics[width = 1\columnwidth]{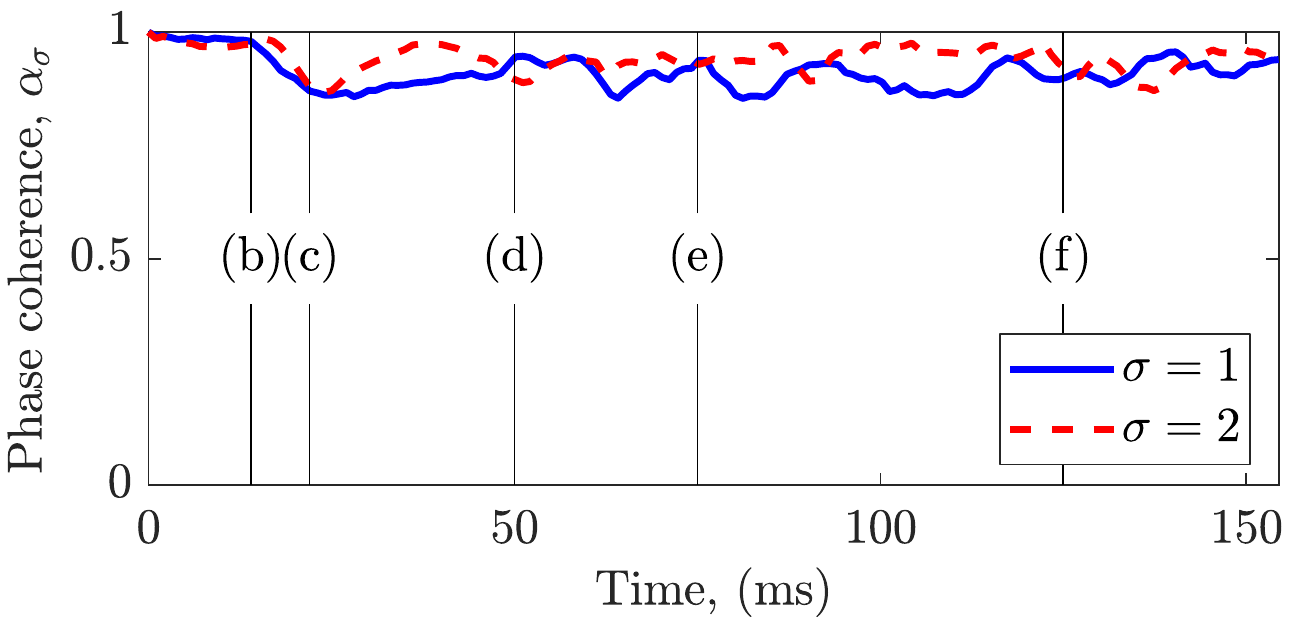}
      \caption{Phase coherence $\alpha_\sigma$ for each component $\sigma=\{1,2\}$ following an instantaneous quench from an unmodulated miscible BEC to a domain supersolid state. Labels in the plot coincide with the frames shown in Fig.~\ref{fig:S2}.}
      \label{fig:S3}
\end{figure}

We can dynamically characterize the supersolid quality by plotting the phase coherence over time. Following Ref.~\cite{Tanzi2019} (see also Refs.~\cite{Bland2022tds,ilzhofer2021phase}) we define the phase coherence as 
\begin{align}
    \alpha_\sigma = 1 - \frac{2}{\pi}\frac{\int_\mathcal{R}\text{d}x\text{d}y\,|\psi_\sigma(x,y)|^2|\theta_\sigma(x,y)-\beta|}{\int_\mathcal{R}\text{d}x\text{d}y\,|\psi_\sigma(x,y)|^2}\,,
\end{align}
where $\theta_\sigma(x,y)$ is the phase of $\psi_\sigma(x,y)$ in the $z=0$ plane, and $\beta$ is a fitting parameter to maximize $\alpha_\sigma$ at each time. The integration region $\mathcal{R}$ encompasses the cloud. From this definition $\alpha_\sigma=1$ corresponds to perfect phase coherence across the BEC. In Fig.~\ref{fig:S3} we present the dynamical evolution of the phase coherence after the instantaneous quench presented in Fig.~\ref{fig:S2}. Throughout the total time evolution, $\alpha_\sigma$ does not go below 0.85 for either component, suggesting excellently maintained phase coherence. 

\end{document}